\documentclass[preprint,prd,aps,showpacs,showkeys,nofootinbib]{revtex4}
\usepackage{amsmath}
\usepackage{graphicx}
\usepackage{color}
\usepackage{xcolor}
\begin{document}


\title{Lepton-flavor violation and muon $(g-2)$ in the flavor-dependent $U(1)_F$ model}

\author{Yu-Ju Peng$^{1,2,3}$, Feng-Yan Niu$^{1,2,3}$\footnote{ffyyniu@163.com}, Qi-Zhen Qin$^{1,2,3}$, Zhan Cao$^{1,2,3}$, Jin-Lei Yang$^{1,2,3}$\footnote{jlyang@hbu.edu.cn}}

\affiliation{Department of Physics, Hebei University, Baoding, 071002, China$^1$\\
	Key Laboratory of High-precision Computation and Application of Quantum Field Theory of Hebei Province, Baoding, 071002, China$^2$\\
	Research Center for Computational Physics of Hebei Province, Baoding, 071002, China$^3$}

\begin{abstract}
Lepton flavor violation (LFV) processes are forbidden in the standard model (SM), hence the observation of LFV transitions would represent a clear signal of new physics beyond the SM. In this work, we investigate the muon anomalous magnetic dipole moments (MDM) and LFV processes $l_{j}^{-}\to l_{i}^{-}\gamma $ and $l_{j}^{-}\to l_{i}^{-}l_{i}^{-}l_{i}^{+}$ in the extension of the SM with $U(1)_F$ local gauge symmetry (FDM). The muon anomalous MDM is one of the most precisely measured quantities in particle physics, which can be used to constrain the contributions from new couplings in the FDM. The newly incorporated $U(1)_F$ charges are correlated with the flavor properties of fermions, hence the newly added Yukawa couplings make contributions to these LFV processes. Considering the latest experimental constraints on the muon anomalous MDM, the new interactions in the FDM can make significant contributions to the LFV processes $l_{j}^{-}\to l_{i}^{-}\gamma $ and $l_{j}^{-}\to l_{i}^{-}l_{i}^{-}l_{i}^{+}$, the predicted branching ratios can well reach the future experimental sensitivity. %

\end{abstract}


\maketitle

\section{Introduction\label{sec1}}
\indent\indent

Lepton-flavor-violation (LFV), if observed in future experiments, is obvious evidence of new physics (NP) beyond the standard model (SM), because the lepton-flavor number is conserved in the SM. Since the processes do not suffer from a large ambiguity due to the hadronic matrix elements, detailed analysis of the LFV processes will reveal some properties of the high-energy physics. In Table~\ref{tab1}, we show the present experimental limits and future sensitivities for the LFV processes $l_j^-\rightarrow l_i^-\gamma,\;l_j^-\rightarrow l_i^-l_i^-l_i^+$~\cite{MEGII:2025gzr,Baldini:2013ke,Aubert:2009ag,Hayasaka:2013dsa,Bellgardt:1987du,Blondel:2013ia,Hayasaka:2010np}.
Several predictions for these LFV processes have been obtained in the framework of various SM extensions~\cite{Ilakovac:1994kj,Diaz:2000cm,Kakizaki:2003jk,Arganda:2005ji,Toma:2013zsa,Zhang:2014osa,Zhao:2015dna,Aebischer:2019blw,Colangelo:2025nbd}. In this work, we analyze these LFV processes in the extension of the SM with $U(1)_F$ local gauge symmetry (FDM). The latest experimental results for the muon anomalous magnetic dipole moment (MDM) show that the measured value is in close agreement with the SM prediction~\cite{Aliberti:2025beg,Boccaletti:2024guq}. This implies that the NP contributions to the muon MDM are tightly constrained by the latest experimental data. Hence, the new contributions in the FDM to the muon MDM are also analyzed in this work.

The fermion sector present in the FDM offers a simultaneous resolution to the puzzles of flavor mixings and fermions mass hierarchy. And utilizing the Type I see-saw mechanism~\cite{Weinberg:1979sa,Hisano:1995nq,Gell-Mann:1979vob,Mohapatra:1979ia}, the model naturally accounts for the generation of non-zero Majorana neutrino masses~\cite{Canetti:2012kh,Abada:2007ux}. In the FDM, the additional $U(1)_F$ charges are related to the flavor of fermions, and as a result, the newly introduced gauge couplings are flavor dependent, which can make significant contributions to the LFV process. In order to realize the proposed mass matrices of fermions, two Higgs doublets and one scalar singlet are introduced, the corresponding Yukawa couplings can make contributions to these LFV processes and the muon anomalous MDM. Hence, the latest experimental results on muon anomalous MDM are considered to limit the relevant couplings. 

The paper is organized as follows. In Sec.~II, the main ingredients of the the FDM are summarized briefly by introducing the scalar potential and the general mass terms. We derive the corresponding analytical expressions through theoretical calculations of the muon anomalous MDM and the decay widethes of LFV processes in Sec.~III. Subsequently, the numerical results predicted in the FDM are presented and analyzed in Sec.~IV. Conslusions are made in Sec.~V, and the tedious formulae are collected in the appendices.

\begin{table*}
\begin{tabular*}{\textwidth}{@{\extracolsep{\fill}}lll@{}}
\hline
LFV process & Present limit & Future sensitivity\\
\hline
$\mu\rightarrow e\gamma$ & $<1.5\times10^{-13}$~\cite{MEGII:2025gzr} & $\sim6\times10^{-14}$~\cite{Baldini:2013ke}\\
$\mu\rightarrow 3e$ & $<1\times10^{-12}$~\cite{Bellgardt:1987du} & $\sim10^{-16}$~\cite{Blondel:2013ia}\\
$\tau\rightarrow e\gamma$ & $<3.3\times10^{-8}$~\cite{Aubert:2009ag} & $\sim10^{-8}-10^{-9}$~\cite{Hayasaka:2013dsa}\\
$\tau\rightarrow 3e$ & $<2.7\times10^{-8}$~\cite{Hayasaka:2010np} & $\sim10^{-9}-10^{-10}$~\cite{Hayasaka:2013dsa}\\
$\tau\rightarrow \mu\gamma$ & $<4.4\times10^{-8}$~\cite{Aubert:2009ag} & $\sim10^{-8}-10^{-9}$~\cite{Hayasaka:2013dsa}\\
$\tau\rightarrow 3\mu$ & $<2.1\times10^{-8}$~\cite{Hayasaka:2010np} & $\sim10^{-9}-10^{-10}$~\cite{Hayasaka:2013dsa}\\
\hline
\end{tabular*}
\caption{Present limits and future sensitivities for the branching ratios for the LFV processes.}
\label{tab1}
\end{table*}

\section{The flavor-dependent model}\label{sec2}
\indent\indent

The gauge group of the FDM is $SU(3)_C\otimes SU(2)_L\otimes U(1)_Y\otimes U(1)_F$, where the additional local gauge group $U(1)_F$ is associated with the flavor of particles~\cite{Yang:2024znv,Yang:2024kfs}. In the FDM, the third generation of fermions obtains masses through the tree-level couplings with the SM scalar doublet, and the first two generations of fermions achieve masses through the tree-level mixings with the third generation. Hence, two additional scalar doublets are introduced in the FDM to realize the tree-level mixings of the first two generations and the third generation. In addition, we implement the Type‑I seesaw mechanism to generate tiny neutrino masses as well as neutrino flavor mixings. Since the $U(1)_F$ charges are associated with the particles’ flavor in the FDM, and the third generation of right-handed neutrinos is a trivial singlet under the gauge group of the model, only two right-handed neutrinos with nonzero $U(1)_F$ charges are introduced naturally. As a consequence, one active neutrino remains massless at tree level, and this is an important feature of the model.

All fields in the FDM and the corresponding gauge symmetry charges are presented in Tab.~\ref{tab2}, where $\Phi_3$ corresponds to the SM Higgs doublet, the nonzero constant $z$ denotes the extra $U(1)_F$ charge. It can be noted in Tab.~\ref{tab2} that there are only two generations of right-handed neutrinos in the FDM, because both $U(1)_F$ and $U(1)_Y$ charges of the third generation of right-handed neutrinos $\nu_{R_3}$ are zero, which is trivial. In addition, it is obvious that the chiral anomaly cancellation can be guaranteed for the fermionic charges presented in Tab.~\ref{tab2}.

\subsection{The scalar sector of the FDM}\label{sec2-1}
\begin{table*}
	\begin{tabular*}{\textwidth}{@{\extracolsep{\fill}}lllll@{}}
		\hline
		Multiplets & $SU(3)_C$ & $SU(2)_L$ & $U(1)_Y$ & $U(1)_F$\\
		\hline
		$l_1=(\nu_{1L},e_{1L})^T$ & 1 & 2 & $-\frac{1}{2}$ & $z$\\
		$l_2=(\nu_{2L},e_{2L})^T$ & 1 & 2 & $-\frac{1}{2}$ & $-z$\\
		$l_3=(\nu_{3L},e_{3L})^T$ & 1 & 2 & $-\frac{1}{2}$ & $0$\\
		$\nu_{1R}$                & 1 & 1 & $0$            & $-z$ \\
		$\nu_{2R}$                & 1 & 1 & $0$            & $z$ \\
		$e_{1R}$                  & 1 & 1 & $-1$           & $-z$ \\
		$e_{2R}$                  & 1 & 1 & $-1$           & $z$ \\
		$e_{3R}$                  & 1 & 1 & $-1$           & $0$ \\
		$q_1=(u_{1L},d_{1L})^T$   & 3 & 2 & $\frac{1}{6}$  & $z$\\
		$q_2=(u_{2L},d_{2L})^T$   & 3 & 2 & $\frac{1}{6}$  & $-z$\\
		$q_3=(u_{3L},d_{3L})^T$   & 3 & 2 & $\frac{1}{6}$  & $0$\\
		$d_{1R}$                  & 3 & 1 & -$\frac{1}{3}$ & $-z$ \\
		$d_{2R}$                  & 3 & 1 & -$\frac{1}{3}$ & $z$ \\
		$d_{3R}$                  & 3 & 1 & -$\frac{1}{3}$ & $0$ \\
		$u_{1R}$                  & 3 & 1 & $\frac{2}{3}$  & $-z$ \\
		$u_{2R}$                  & 3 & 1 & $\frac{2}{3}$  & $z$ \\
		$u_{3R}$                  & 3 & 1 & $\frac{2}{3}$  & $0$ \\
		$\Phi_1=(\phi_1^{+},\phi_1^{0})^T$   & 1 & 2 & $\frac{1}{2}$  & $z$\\
		$\Phi_2=(\phi_2^{+},\phi_2^{0})^T$   & 1 & 2 & $\frac{1}{2}$  & $-z$\\
		$\Phi_3=(\phi_3^{+},\phi_3^{0})^T$   & 1 & 2 & $\frac{1}{2}$  & 0\\
		$\chi$   & 1 & 1 & 0  & $2z$\\
		\hline
	\end{tabular*}
	\caption{Matter content in the FDM, where the nonzero constant $z$ denotes the extra $U(1)_F$ charge.}
	\label{tab2}
\end{table*}

The scalar sector is extended by introducing two new doublets and one new singlet~\cite{Yang:2024znv,Cao:2025zwn,Yang:2024kfs,Yang:2024duo}
\begin{eqnarray}
	&&\Phi_1=\left(\begin{array}{c}\phi_1^+\\ \frac{1}{\sqrt2}(i A_1+S_1+v_1)\end{array}\right)\sim(2,\frac{1}{2},z),\nonumber\\
	&&\Phi_2=\left(\begin{array}{c}\phi_2^+\\ \frac{1}{\sqrt2}(i A_2+S_2+v_2)\end{array}\right)\sim(2,\frac{1}{2},-z),\nonumber\\
	&&\Phi_3=\left(\begin{array}{c}\phi_3^+\\ \frac{1}{\sqrt2}(i A_3+S_3+v_3)\end{array}\right)\sim(2,\frac{1}{2},0),\nonumber\\
	&&\chi=\frac{1}{\sqrt2}(i A_{\chi}+S_{\chi}+v_\chi)\sim(1,0,2z),\label{eq2}
\end{eqnarray}
and $v_i\;(i=1,\;2,\;3),\;v_\chi$ are the VEVs of $\Phi_i,\;\chi$ respectively, where $(v_1^2+v_2^2+v_3^2)^{1/2}= v\approx 246\;{\rm GeV}$.

The scalar potential in the FDM can be written as
\begin{eqnarray}
	&&V=-M_{\Phi_1}^2 \Phi_1^\dagger\Phi_1-M_{\Phi_2}^2 \Phi_2^\dagger\Phi_2-M_{\Phi_3}^2 \Phi_3^\dagger\Phi_3-M_{\chi}^2\chi^*\chi+\lambda_{\chi} (\chi^*\chi)^2+\lambda_1 (\Phi_1^\dagger\Phi_1)^2\nonumber\\
	&&\qquad+\lambda_2 (\Phi_2^\dagger\Phi_2)^2+\lambda_3 (\Phi_3^\dagger\Phi_3)^2+\lambda'_4 (\Phi_1^\dagger\Phi_1)(\Phi_2^\dagger\Phi_2)+\lambda_4'' (\Phi_1^\dagger\Phi_2)(\Phi_2^\dagger\Phi_1)\nonumber\\
	&&\qquad+\lambda_5' (\Phi_1^\dagger\Phi_1)(\Phi_3^\dagger\Phi_3)+\lambda_5'' (\Phi_1^\dagger\Phi_3)(\Phi_3^\dagger\Phi_1)+\lambda_6' (\Phi_2^\dagger\Phi_2)(\Phi_3^\dagger\Phi_3)+\lambda_6'' (\Phi_2^\dagger\Phi_3)(\Phi_3^\dagger\Phi_2)\nonumber\\
	&&\qquad+\lambda_7 (\Phi_1^\dagger\Phi_1)(\chi^*\chi)+\lambda_{8} (\Phi_2^\dagger\Phi_2)(\chi^*\chi)+\lambda_{9} (\Phi_3^\dagger\Phi_3)(\chi^*\chi)+[\lambda_{10} (\Phi_3^\dagger\Phi_1)(\Phi_3^\dagger\Phi_2)\nonumber\\
	&&\qquad+\kappa(\Phi_1^\dagger\Phi_2)\chi+h.c.],\label{eq3}
\end{eqnarray}

Based on the scalar potential in Eq.~(\ref{eq3}), the tadpole equations in the FDM can be written as
\begin{eqnarray}
	&&M_{\Phi_1}^2=\lambda_1v_1^2+\frac{1}{2}\Big[(\lambda_4'+\lambda_4'') v_2^2+(\lambda_5'+\lambda_5'') v_3^2+\frac{v_2}{v_1}v_3^2 {\rm Re}(\lambda_{10})+\sqrt2\frac{v_2}{v_1} v_\chi {\rm Re}(\kappa)+\lambda_7v_\chi^2\Big],\nonumber\\
	&&M_{\Phi_2}^2=\lambda_2v_2^2+\frac{1}{2}\Big[(\lambda_4'+\lambda_4'') v_1^2+(\lambda_6'+\lambda_6'') v_3^2+\frac{v_1}{v_2}v_3^2 {\rm Re}(\lambda_{10})+\sqrt2\frac{v_1}{v_2} v_\chi {\rm Re}(\kappa)+\lambda_8v_\chi^2\Big],\nonumber\\
	&&M_{\Phi_3}^2=\lambda_3v_3^2+{\rm Re}(\lambda_{10})v_1v_2+\frac{1}{2}[(\lambda_5'+\lambda_5'') v_1^2+(\lambda_6'+\lambda_6'') v_2^2+\lambda_9 v_c^2],\nonumber\\
	&&M_{\chi}^2=\lambda_\chi v_\chi^2+\frac{1}{2}\Big[\lambda_7 v_1^2+\lambda_8 v_2^2+\lambda_9 v_3^2+\sqrt2\frac{v_1v_2}{v_\chi}  {\rm Re}(\kappa)\Big].\label{eq4}
\end{eqnarray}

On the basis $(S_1,\;S_2,\;S_3,\;S_\chi)$, the CP-even Higgs squared mass matrix in the FDM is
\begin{eqnarray}
	&&M_{h}^2=\left(\begin{array}{*{20}{cccc}}
		M_{h,11}^2 & M_{h,12}^2 & M_{h,13}^2 & M_{h,14}^2 \\ [6pt]
		M_{h,12}^2 & M_{h,22}^2 & M_{h,23}^2 & M_{h,24}^2 \\ [6pt]
		M_{h,13}^2 & M_{h,23}^2 & M_{h,33}^2 & M_{h,34}^2 \\ [6pt]
		M_{h,14}^2 & M_{h,24}^2 & M_{h,34}^2 & M_{h,44}^2 \\ [6pt]
	\end{array}\right),
\end{eqnarray}
where
\begin{eqnarray}
	&&M_{h,11}^2=2\lambda_1v_1^2-\frac{v_2}{2v_1}\Big[v_3^2 {\rm Re}(\lambda_{10})+\sqrt2 v_\chi {\rm Re}(\kappa)\Big],\nonumber\\
	&&M_{h,22}^2=2\lambda_2v_2^2-\frac{v_1}{2v_2}\Big[v_3^2 {\rm Re}(\lambda_{10})+\sqrt2 v_\chi {\rm Re}(\kappa)\Big],\nonumber\\
	&&M_{h,33}^2=2\lambda_3v_3^2,\;\;M_{h,44}^2=2\lambda_\chi v_\chi^2-\frac{\sqrt2v_1v_2}{2v_\chi}  {\rm Re}(\kappa),\nonumber\\
	&&M_{h,12}^2=(\lambda_4'+\lambda_4'') v_1v_2+\frac{1}{2}{\rm Re}(\lambda_{10})v_3^2+\frac{\sqrt2}{2}{\rm Re}(\kappa)v_\chi,\nonumber\\
	&&M_{h,13}^2=(\lambda_5'+\lambda_5'')v_1v_3+{\rm Re}(\lambda_{10})v_2v_3,\;\;M_{h,14}^2=\lambda_7v_1v_\chi+\frac{\sqrt2}{2}{\rm Re}(\kappa)v_2,\nonumber\\
	&&M_{h,23}^2=(\lambda_6'+\lambda_6'')v_2v_3+{\rm Re}(\lambda_{10})v_1v_3,\;\;M_{h,24}^2=\lambda_8v_2v_\chi+\frac{\sqrt2}{2}{\rm Re}(\kappa)v_1,\nonumber\\
	&&M_{h,34}^2=\lambda_9v_3v_\chi.\label{eqmh}
\end{eqnarray}

The tadpole equations in Eq.~(\ref{eq4}) are used to obtain the matrix elements above. Then, on the basis $(A_1,\;A_2,\;A_3,\;A_\chi)$, the squared mass matrix of CP-odd Higgs in the FDM can be written as
\begin{eqnarray}
	&&M_{A}^2=\left(\begin{array}{*{20}{cccc}}
		M_{A,11}^2 & M_{A,12}^2 & M_{A,13}^2 & M_{A,14}^2 \\ [6pt]
		M_{A,12}^2 & M_{A,22}^2 & M_{A,23}^2 & M_{A,24}^2 \\ [6pt]
		M_{A,13}^2 & M_{A,23}^2 & M_{A,33}^2 & M_{A,34}^2 \\ [6pt]
		M_{A,14}^2 & M_{A,24}^2 & M_{A,34}^2 & M_{A,44}^2 \\ [6pt]
	\end{array}\right),
\end{eqnarray}
where
\begin{eqnarray}
	&&M_{A,11}^2=-\frac{v_2}{2v_1}[{\rm Re}(\lambda_{10})v_3^2+\sqrt2 v_\chi {\rm Re} (\kappa)],\;\;M_{A,33}^2=-2{\rm Re}(\lambda_{10})v_1v_2,\nonumber\\
	&&M_{A,22}^2=-\frac{v_1}{2v_2}[{\rm Re}(\lambda_{10})v_3^2+\sqrt2 v_\chi {\rm Re} (\kappa)],\;\;M_{A,44}^2=-\frac{\sqrt2 v_1v_2}{2v_\chi}{\rm Re} (\kappa),\nonumber\\
	&&M_{A,12}^2=\frac{\sqrt 2}{2} v_\chi {\rm Re} (\kappa)-\frac{1}{2}{\rm Re}(\lambda_{10})v_3^2,\;\;M_{A,13}^2={\rm Re}(\lambda_{10})v_2v_3,\nonumber\\
	&&M_{A,14}^2=\frac{\sqrt 2}{2} v_2 {\rm Re} (\kappa),\;\;M_{A,23}^2={\rm Re}(\lambda_{10})v_1v_3,\;\;M_{A,24}^2=-\frac{\sqrt 2}{2} v_1 {\rm Re} (\kappa),\nonumber\\
	&&M_{A,34}^2=0.\label{eqmA}
\end{eqnarray}

On the basis $(\phi_1^+,\;\phi_2^+,\;\phi_3^+)$ and $(\phi_1^-,\;\phi_2^-,\;\phi_3^-)^T$, the squared mass matrix of singly charged Higgs in the FDM can be written as
\begin{eqnarray}
	&&M_{H^\pm}^2=\left(\begin{array}{*{20}{ccc}}
		M_{H^\pm,11}^2 & M_{H^\pm,12}^2 & M_{H^\pm,13}^2 \\ [6pt]
		(M_{H^\pm,12}^2)^* & M_{H^\pm,22}^2 & M_{H^\pm,23}^2 \\ [6pt]
		(M_{H^\pm,13}^2)^* & (M_{H^\pm,23}^2)^* & M_{H^\pm,33}^2 \\ [6pt]
	\end{array}\right),
\end{eqnarray}
where
\begin{eqnarray}
	&&M_{H^\pm,11}^2=-\frac{v_2}{2v_1}[{\rm Re}(\lambda_{10})v_3^2+\sqrt2 v_\chi {\rm Re} (\kappa)]-\frac{1}{2}(\lambda_4'' v_2^2+\lambda_5'' v_3^2),\nonumber\\
	&&M_{H^\pm,22}^2=-\frac{v_1}{2v_2}[{\rm Re}(\lambda_{10})v_3^2+\sqrt2 v_\chi {\rm Re} (\kappa)]-\frac{1}{2}(\lambda_4'' v_1^2+\lambda_6'' v_3^2),\nonumber\\
	&&M_{H^\pm,33}^2=-{\rm Re}(\lambda_{10})v_1v_2-\frac{1}{2}(\lambda_5'' v_1^2+\lambda_6'' v_2^2),\nonumber\\
	&&M_{H^\pm,12}^2=\frac{\sqrt 2}{2}v_\chi\kappa+\frac{1}{2}\lambda_4'' v_1 v_2,\;\;M_{H^\pm,13}^2=\frac{1}{2}v_3(\lambda_5'' v_1+\lambda_{10}^*v_2),\nonumber\\
	&&M_{H^\pm,23}^2=\frac{1}{2}v_3(\lambda_6'' v_2+\lambda_{10}^*v_1).\label{eqmCH}
\end{eqnarray}

It is easy to verify that there are two neutral Goldstones and one singly charged Goldstone in the FDM.

\subsection{The fermion masses in the FDM}\label{sec2-2}

Based on the matter content listed in Tab.~\ref{tab2}, the Yukawa couplings in the FDM can be written as
\begin{eqnarray}
	&&\mathcal{L}_Y=Y_u^{33}\bar q_3 \tilde \Phi_3 u_{R_3}+Y_d^{33}\bar q_3 \Phi_3 d_{R_3}+Y_u^{32}\bar q_3 \tilde{\Phi}_1 u_{R_2}+Y_u^{23}\bar q_2 \tilde \Phi_1 u_{R_3}+Y_d^{32}\bar q_3 \Phi_2 d_{R_2}\nonumber\\
	&&\qquad\; +Y_d^{23}\bar q_2 \Phi_2 d_{R_3}+Y_u^{21}\bar q_2 \tilde{\Phi}_3 u_{R_1}+Y_u^{12}\bar q_1 \tilde \Phi_3 u_{R_2}+Y_d^{21}\bar q_2 \Phi_3 d_{R_1}+ Y_d^{12}\bar q_1 \Phi_3 d_{R_2}\nonumber\\
	&&\qquad\; +Y_u^{31}\bar q_3 \tilde{\Phi}_2 u_{R_1}+Y_u^{13}\bar q_1 \tilde \Phi_2 u_{R_3}+Y_d^{31}\bar q_3 \Phi_1 d_{R_1}+Y_d^{13}\bar q_1 \Phi_1 d_{R_3}\nonumber\\
	&&\qquad\; +Y_e^{33}\bar l_3 \Phi_3 e_{R_3}+Y_e^{32}\bar l_3 \Phi_2 e_{R_2}+Y_e^{23}\bar l_2 \Phi_2 e_{R_3}+Y_e^{21}\bar l_2 \Phi_3 e_{R_1}+ Y_e^{12}\bar l_1 \Phi_3 e_{R_2}\nonumber\\
	&&\qquad\; +Y_e^{31}\bar l_3 \Phi_1 e_{R_1}+Y_e^{13}\bar l_1 \Phi_1 e_{R_3}+Y_R^{11}\bar\nu^c_{R_1}\nu_{R_1}\chi+Y_R^{22}\bar\nu^c_{R_2}\nu_{R_2} \chi^*+Y_D^{21}\bar l_2 \tilde \Phi_3 \nu_{R_1}\nonumber\\
	&&\qquad\; +Y_D^{12}\bar l_1 \tilde \Phi_3 \nu_{R_2}+Y_D^{31}\bar l_3 \tilde \Phi_2 \nu_{R_1}+Y_D^{32}\bar l_3 \tilde \Phi_1 \nu_{R_2}+M_{R,12} \bar\nu^c_{R_1}\nu_{R_2}+h.c.,\label{eq11}
\end{eqnarray}

Then the mass matrices of quarks and leptons can be written as
\begin{eqnarray}
	&&m_q=\left(\begin{array}{ccc} 0 & m_{q,12} & m_{q,13}\\
		m_{q,12}^* & 0 & m_{q,23}\\
		m_{q,13}^* & m_{q,23}^* & m_{q,33}\end{array}\right),m_e=\left(\begin{array}{ccc} 0 & m_{e,12} & m_{e,13}\\
		m_{e,12}^* & 0 & m_{e,23}\\
		m_{e,13}^* & m_{e,23}^* & m_{e,33}\end{array}\right),m_\nu=\left(\begin{array}{cc} 0 & M_D^T\\
		M_D & M_R\end{array}\right),\label{eq12}
\end{eqnarray}
where $q=u,d$, the parameters $m_{q,33}$ and $m_{e,33}$ are real, $M_D$ is $2\times3$ Dirac mass matrix and $M_R$ is $2\times2$ Majorana mass matrix (the nonzero neutrino masses are obtained by the Type I see-saw mechanism). Then considering the measured fermionic masses, under the approximation $|m_{f,12}|,\;|m_{f,13}|,\;|m_{f,23}|\ll |m_{f,33}|\;(f=u,\;d,\;e)$ one can obtain
\begin{eqnarray}
	&&m_{f,33}=m_{f_3}-m_{f_1}-m_{f_2},\nonumber\\
	&&|m_{f,23}|=[(m_{f_1}+m_{f_2})m_{f,33}-|m_{f,13}|^2]^{1/2},\nonumber\\
	&&|m_{f,12}|=\frac{|m_{f,13}||m_{f,23}|}{|m_{f,33}|} \cos(\theta_{f,12}+\theta_{f,23}-\theta_{f,13})\nonumber\\
	&&\qquad\qquad+\Big\{[\frac{|m_{f,13}||m_{f,23}|}{|m_{f,33}|^2}\cos(\theta_{f,12}+\theta_{f,23}-\theta_{f,13})]^2+\frac{m_{f_1}m_{f_2}}{|m_{f,33}|^2}\Big\}^{1/2}|m_{f,33}|,\label{eq13}
\end{eqnarray}
where $\theta_{f,ij}\;(ij=12,\;13,\;23)$ are defined as $m_{f,ij}=|m_{f,ij}|e^{i\theta_{f,ij}}$, and $m_{f_k}\;(k=1,\;2,\;3)$ is $k-$generation fermion $f$ mass. In the numerical computations, we take the measured fermion masses as input to fix $m_{f,33},\;|m_{f,23}|,\;|m_{f,12}|$ by Eq.~\eqref{eq13}. Then the predicted fermion masses can well fit the observations. According to our previous analysis on the neutrino sector of the FDM~\cite{Yang:2024duo}, the differences of squared neutrino masses and neutrino mixing matrix elements can be well fitted by choosing appropriate values of the neutrino mass matrix elements, which affect the numerical results obtained in this work negligibly.

The elements of the matrices in Eq.~(\ref{eq12}) are
\begin{eqnarray}
	&&m_{u,11}=m_{u,22}=0,\;m_{u,33}=\frac{1}{\sqrt2}Y_u^{33}v_3,\;m_{u,12}=\frac{1}{\sqrt2}Y_u^{12}v_3,\;m_{u,13}=\frac{1}{\sqrt2}Y_u^{13}v_1,\nonumber\\
	&&m_{u,23}=\frac{1}{\sqrt2}Y_u^{23}v_2,\label{eqmu}\\
	&&m_{d,11}=m_{d,22}=0,\;m_{d,33}=\frac{1}{\sqrt2}Y_d^{33}v_3,\;m_{d,12}=\frac{1}{\sqrt2}Y_d^{12}v_3,\;m_{d,13}=\frac{1}{\sqrt2}Y_d^{13}v_1,\nonumber\\
	&&m_{d,23}=\frac{1}{\sqrt2}Y_d^{23}v_2,\label{eqmd}\\
	&&m_{e,11}=m_{e,22}=0,\;m_{e,33}=\frac{1}{\sqrt2}Y_e^{33}v_3,\;m_{e,12}=\frac{1}{\sqrt2}Y_e^{12}v_3,\;m_{e,13}=\frac{1}{\sqrt2}Y_e^{13}v_1,\nonumber\\
	&&m_{e,23}=\frac{1}{\sqrt2}Y_e^{23}v_2,\label{eqme}\\
	&&M_{D,11}=M_{D,22}=0,\;\;M_{D,12}=\frac{1}{\sqrt2}Y_D^{12}v_3,\;M_{D,31}=\frac{1}{\sqrt2}Y_D^{31}v_1,\;M_{D,32}=\frac{1}{\sqrt2}Y_D^{32}v_2,\nonumber\\
	&&M_{R,11}=\frac{1}{\sqrt2}Y_R^{11}v_\chi,\;M_{R,22}=\frac{1}{\sqrt2}Y_R^{22}v_\chi.
\end{eqnarray}
In the numerical computations, we take $M_{R,12}=0$ for simplicity, because their effects are highly suppressed by the tiny heavy-light neutrino mixings.

\subsection{The gauge sector of the FDM}\label{sec2-3}

Due to the introducing of an extra $U(1)_F$ local gauge group in the FDM, the covariant derivative corresponding to $SU(2)_L\otimes U(1)_Y\otimes U(1)_F$ is defined as
\begin{eqnarray}
	&&D_\mu=\partial_\mu+i g_2 T_j A_{j\mu}+i g_1 Y B_\mu+i g_{_F} F B'_\mu+i g_{_{YF}} Y B'_\mu,\;(j=1,\;2,\;3),\label{eqCD}
\end{eqnarray}
where $(g_2,\;g_1,\; g_{_F})$, $(T_j,\;Y,\;F)$, $(A_{j\mu},\;B_\mu,\; B'_\mu)$ denote the gauge coupling constants, generators and gauge bosons of groups $(SU(2)_L,\;U(1)_Y,\;U(1)_F)$ respectively, $g_{_{YF}}$ is the gauge coupling constant that arises from the gauge kinetic mixing effect which presents in the models with two Abelian groups. Then the $W$ boson mass can be written as
\begin{eqnarray}
	&&M_W=\frac{1}{2}g_2 (v_1^2+v_2^2+v_3^2)^{1/2},
\end{eqnarray}
where $(v_1^2+v_2^2+v_3^2)^{1/2}=v\approx246\;{\rm GeV}$ and we have $v_1,\;v_2 < v_3$ in the FDM. The $\gamma$, $Z$ and $Z'$ boson masses in the FDM can be written as
\begin{eqnarray}
	&&M_\gamma=0,\;M_Z\approx\frac{1}{2}(g_1^2+g_2^2)^{1/2} v,\;M_{Z'}\approx 2|zg_{_F}| v_\chi,\label{eq20}
\end{eqnarray}
and
\begin{eqnarray}
	&&\gamma=c_W B+s_W A_3,\;Z=-s_W B+c_W A_3+s'_W B',\;Z'=s_W'(s_WB-c_W A_3)+c'_W B'.
\end{eqnarray}
Here, $\gamma,\;Z,\;Z'$ are the mass eigenstates, $z=1$ is adopted in this work, $c_W\equiv \cos \theta_W,\;s_W\equiv \sin \theta_W$ with $\theta_W$ denoting the Weinberg angle. Similarly, We define $s_W'\equiv \sin \theta'_W$, $c_W'\equiv \cos \theta'_W$, where $\theta_W'$ represents the $Z-Z'$ mixing effect, and the mixing angle $\theta'_W$ can be approximated as 
\begin{equation}
 \theta'_{W}\simeq\frac{M_Z^2}{M_{Z'}^2}.
\end{equation}
The approximation $M_{Z}\ll M_{Z'}$ is applied to obtain the above equation.

\section{Lepton flavor violation and $(g-2)_\mu$ in the FDM\label{sec3}}
\indent\indent

The new interactions introduced in the FDM can make significant contributions to the physical observations considered in this work. The analytical calculations of $\Delta a_\mu$ and the LFV processes at the one-loop level are presented in this section.

\subsection{$(g-2)_\mu$}
The anomalous MDM of lepton $a_l$~\cite{Schwinger:1948iu} is one of the most precisely measured and calculated quantities in elementary particle physics, which also provides one of the strongest tests of the SM.  Several predictions for the muon MDM have been discussed in the framework of various SM extensions~\cite{Bennett:2006fi,Mohr:2008fa,Abel:1991dv,Moroi:1995yh,Feng:2001tr,Martin:2001st,Diaz:2002tp,Cheung:2009fc,Zhao:2014dxa,Feng:2008cn,Feng:2008nm,Feng:2009gn,Yang:2009zzh,
Padley:2015uma,Li:2018aov,Li:2020dbg,Cao:2021lmj,Chen:2021rnl,Yin:2021yqy,Yin:2020afe,Sabatta:2019nfg,
vonBuddenbrock:2019ajh,vonBuddenbrock:2016rmr,Okada:2016wlm,Fukuyama:2016mqb,Belanger:2017vpq,Megias:2017dzd,Tran:2018kxv,
g-2muonQCD,g-2muon,g-2muon1,g-2muon2,g-2muon3,g-2muon4,g-2muon5,g-2muon6,g-2muon7,g-2muon8,g-2muon9,g-2muon10,
g-2muon11,g-2muon12,g-2muon13,g-2muon14,g-2muon15,g-2muon16,g-2muon17,g-2muon18,g-2muon19,g-2muon20,
g-2muon21,g-2muon22,g-2muon23,g-2muon24,g-2muon25,g-2muon26,g-2muon27,g-2muon28,g-2muon29,g-2muon30}.
Recently, Fermilab reports their latest result of the muon anomalous MDM, the new experimental average for the difference between the experimentally measured value and the SM prediction reads~\cite{Aliberti:2025beg}
\begin{eqnarray}
&&\Delta a_\mu=a_\mu^{exp}-a_\mu^{SM}=26(66)\times10^{-11}.
\label{eqamu}
\end{eqnarray}
It indicates that the tension between experiment and the SM prediction is increased to $0.5$ standard deviations, significantly below the 5 standard deviations discovery threshold. This latest result implies strict constraints on the NP contributions to the muon MDM, which can actually be expressed as~\cite{Moroi:1995yh}
\begin{eqnarray}
	{\cal L}_{\rm MDM} = \frac{e}{4m_\mu} F_2
	\bar{\mu} \sigma_{\rho\lambda} \mu F^{\rho\lambda}.
	\label{mag_mom}
\end{eqnarray}
Here, $e$ is the electric charge, $m_\mu$ is the muon mass,
$\sigma_{\rho\lambda}=\frac{i}{2}[\gamma_\rho ,\gamma_\lambda]$,
$F_{\rho\lambda}$ denotes the field strength of the photon field and $F_2$ is the magnetic form factor. Then $a_\mu$ can be expressed as~\cite{Moroi:1995yh}
\begin{eqnarray}
	a_\mu = F_2.
\end{eqnarray}

In the FDM, the Feynman diagrams making dominant contributions to the muon anomalous MDM $\Delta a_\mu$ are depicted in Fig.~\ref{figllr} with $i=j=2$, where $\nu_{k}$ denotes neutrino, $H^\pm_{l}$ is charged Higgs boson, $l_k^-$ denotes the charged lepton, $h_{\alpha} $ and $A_{\alpha}$ denote scalar and pseudoscalar respectively. Then the contributions from Fig.~\ref{figllr} to the muon anomalous MDM can be written as
\begin{eqnarray}
&&\Delta a_\mu = \Delta a_\mu^{(a)} + \Delta a_\mu^{(b)},
\label{oneloop MDM}
\end{eqnarray}
where~\cite{Moroi:1995yh}
\begin{eqnarray}
\Delta a_\mu^{(a)} &&=m_\mu \sum_{k,l} \Big\{-m_\mu (C_{\bar l_2\nu_k H_l^\pm}^LC_{\bar l_2\nu_k H_l^\pm}^L + C_{\bar l_2\nu_k H_l^\pm}^RC_{\bar l_2\nu_k H_l^\pm}^R) m_{H^\pm_{l}}^2
J_5(m_{\nu_{k}}^2,m_{H^\pm_{l}}^2,m_{H^\pm_{l}}^2
\nonumber\\&&
,m_{H^\pm_{l}}^2,m_{H^\pm_{l}}^2)+ m_{\nu_{k}}C_{\bar l_2\nu_k H_l^\pm}^LC_{\bar l_2\nu_k H_l^\pm}^R
J_4(m_{\nu_{k}}^2,m_{\nu_{k}}^2,m_{H^\pm_{l}}^2,m_{H^\pm_{l}}^2)\Big\},\label{g-2_nAX}\\
\Delta a_\mu^{(b)} &&=m_\mu \sum_{k,l}\Big[ m_\mu (C_{\bar l_2l_k h_l}^L C_{\bar l_2l_k h_l}^L +C_{\bar l_2l_k h_l}^RC_{\bar l_2l_k h_l}^R) \{ J_4(m_{l_k^-}^2,m_{l_k^-}^2,m_{l_k^-}^2,m_{l_k^-}^2)
\nonumber\\&&
+ m_{h_{\alpha}}^2 J_5(m_{l_k^-}^2,m_{l_k^-}^2,m_{l_k^-}^2,m_{l_k^-}^2,m_{h_{\alpha}}^2)
- J_4(m_{l_k^-}^2,m_{l_k^-}^2,m_{l_k^-}^2,m_{h_{\alpha}}^2) \}
\nonumber\\&&
-2m_{l_k^-} C_{\bar l_2l_k h_l}^LC_{\bar l_2l_k h_l}^R J_4 (m_{l_k^-}^2,m_{l_k^-}^2,m_{l_k^-}^2
,m_{h_{\alpha}}^2) \Big].
\label{g-2_cAX}
\end{eqnarray}
Here, $x_1 =m_{\nu_{k}}^2/m_{H^\pm_{l}}^2$, $x_2=m_{l_k^-}^2/m_{h_{\alpha}}^2,\;x_3=m_{l_k^-}^2/m_{A_{\alpha}}^2,\;(k=1,\;2,\;3)$, and $\Delta a_\mu^{(a,b)}$ correspond to the contributions from Fig.~\ref{figllr} (a), (b) respectively. $C_{abc}^{L,R}$ denotes the constant parts of the interaction vertex about $abc$, and $a, b, c$ denote the particles appearing in the interaction vertex. Explicit expressions for $C_{abc}^{L,R}$ are collected in the Appendix~\ref{wilsonllr}. The loop integral functions $J_4,\; J_5$ can be found in the Appendix~\ref{au}. 

The electron anomalous magnetic moment, $\Delta a_e$, receives contributions from the same one-loop topologies as $\Delta a_\mu$. Since the relevant electron Yukawa couplings are smaller than the corresponding muon couplings, the predicted $\Delta a_e$ is strongly suppressed relative to $\Delta a_\mu$. Moreover, the diagrams with charged scalar at loop can also make contributions to the electron electric dipole moment (EDM), but the contributions are proportional to the neutrino coupling matrix $Y_D$, which are highly suppressed by the tiny neutrino masses.

\subsection{Rare decay $l_j^-\rightarrow l_i^-\gamma$}

\begin{figure}
\setlength{\unitlength}{1mm}
\centering
\includegraphics[width=5in]{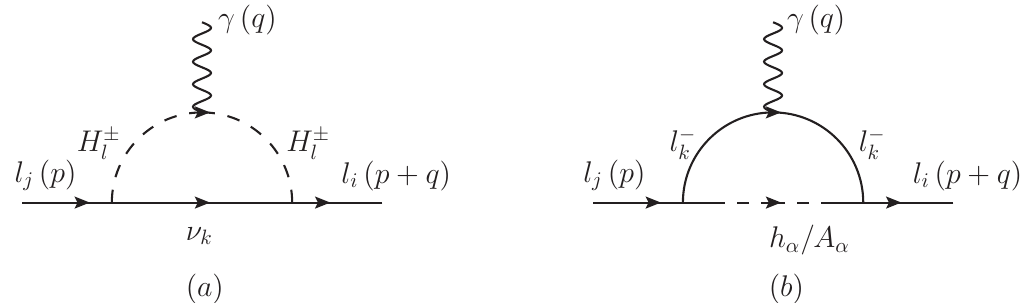}
\vspace{0cm}
\caption[]{Feynman diagrams for the process $l_j^-\rightarrow l_i^-\gamma$. (a) represents the contributions from neutral fermions $\nu_{k} $ and charge scalars loops $H^\pm_{l}$, and (b) represents the contributions from charged fermions $l_k^-$ and neutral scalars $h_{\alpha} $ or pseudoscalars $A_{\alpha}$ loops.}
\label{figllr}
\end{figure}

The off-shell amplitude for $l_j^-\rightarrow l_i^-\gamma$ is generally written as~\cite{Hisano:1995cp,Huang:2024ozb,Yang:2018guw}
\begin{eqnarray}
&&T=e\epsilon^\mu \bar u_i(p+q)[q^2 \gamma_\mu(A_1^LP_L+A_1^RP_R)+m_{l_j}i\sigma_{\mu\nu}q^\nu(A_2^LP_L+A_2^RP_R)]u_j(p),\label{Allr}
\end{eqnarray}
In addition, $p$ is the momentum of the particle $l_j$, $\epsilon$ is the photon polarization vector. Then, the Feynman diagrams contributing to the above amplitude are shown in Fig.~\ref{figllr} with $j>i$.\footnote{Since the model generates dipole, photon-penguin and $Z$/$Z'$-penguin, all of which can contribute to coherent $\mu\to e$ conversion in nuclei, which may play a complementary role to $\mu\to e\gamma$. A complete nuclear-rate calculation is outside the scope of the present paper, we will focus on the study of $\mu\to e$ conversion in the future work.}
The coefficients $A_{1,2}^{L,R}$ in Eq.(\ref{Allr}) can be written as
\begin{eqnarray}
&&A_1^{L,R}=A_1^{(a)L,R}+A_1^{(b)L,R},\nonumber\\
&&A_2^{L,R}=A_2^{(a)L,R}+A_2^{(b)L,R},
\end{eqnarray}
where the concrete expressions for $A_{1,2}^{(a)L,R}$, $A_{1,2}^{(b)L,R}$ corresponding to Fig.~\ref{figllr} (a), (b) can be found in Appendix \ref{wilsonllr}.

Using the amplitude Eq.~(\ref{Allr}), the decay widths for $l_j^-\rightarrow l_i^-\gamma$ can be obtained
\begin{eqnarray}
&&\Gamma(l_j^-\rightarrow l_i^-\gamma)=\frac{e^2}{16\pi}m_{l_j}^5(|A_2^L|^2+|A_2^R|^2).
\end{eqnarray}
The branching ratio then is calculated as
\begin{eqnarray}
&&Br(l_j^-\rightarrow l_i^-\gamma)=\frac{\Gamma(l_j^-\rightarrow l_i^-\gamma)}{\Gamma_{l_j^-}},
\end{eqnarray}
where $\Gamma_{l_j^-}$ is the total decay width of the lepton $l_j^-$. In the numerical calculations, we use $\Gamma_\mu\approx2.996\times10^{-19}{\rm GeV}$ for the muon and $\Gamma_\tau\approx2.265\times10^{-12}{\rm GeV}$ for the tau lepton~\cite{PDG}.

\subsection{Rare decay $l_j^-\rightarrow l_i^-l_i^-l_i^+$}
\begin{figure}
\setlength{\unitlength}{2mm}
\centering
\includegraphics[width=2in]{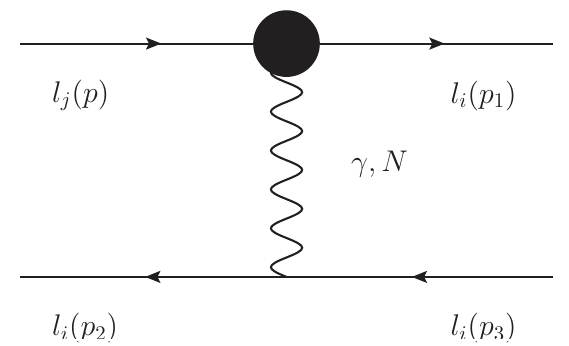}
\vspace{0cm}
\caption[]{Penguin-type diagram for the process $l_j^-\rightarrow l_i^-l_i^-l_i^+$. The black dot indicates an $l_j^-l_i^-\gamma$ vertex such as Fig.~\ref{figllr} or $l_j^-l_i^-N$ vertex where $N$ denotes $Z$ and $Z'$ bosons.}
\label{figl3lPenguin}
\end{figure}
For the process $l_j^-\rightarrow l_i^-l_i^-l_i^+$, the dominant contributions come from penguin-type and box-type diagrams. The contributions from $\gamma$-penguin diagrams can be written as~\cite{Hisano:1995cp}
\begin{eqnarray}
&&T_{\gamma-{\rm penguin}}=\bar u_i(p_1)[q^2 \gamma_\mu(A_1^LP_L+A_1^RP_R)+m_{l_j}i\sigma_{\mu\nu}q^\nu(A_2^LP_L+A_2^RP_R)]u_j(p)\nonumber\\
&&\qquad\qquad\quad\times\frac{e^2}{q^2}\bar u_i(p_2)\gamma^\mu \nu_i(p_3)-(p_1\leftrightarrow p_2),\label{rpenguin}
\end{eqnarray}
and the ones from $N$-penguin (here $N$ represents $Z$ and $Z'$ bosons) diagrams can be written as~\cite{Hisano:1995cp}
\begin{eqnarray}
&&T_{N-{\rm penguin}}=\frac{e^2}{m_N^2}\bar u_i(p_1)\gamma_\mu(F^LP_L+F^RP_R)u_j(p)\bar u_i(p_2)\gamma^\mu(C_{\bar l_iNl_i}^LP_L\nonumber\\
&&\qquad\qquad\quad+C_{\bar l_iNl_i}^RP_R)\nu_i(p_3)-(p_1\leftrightarrow p_2).\label{Npenguin}
\end{eqnarray}
The concrete expressions for $F^{L,R}$ are given in Appendix~\ref{wilsonl3l}.

\begin{figure}
\setlength{\unitlength}{1mm}
\centering
\includegraphics[width=4in]{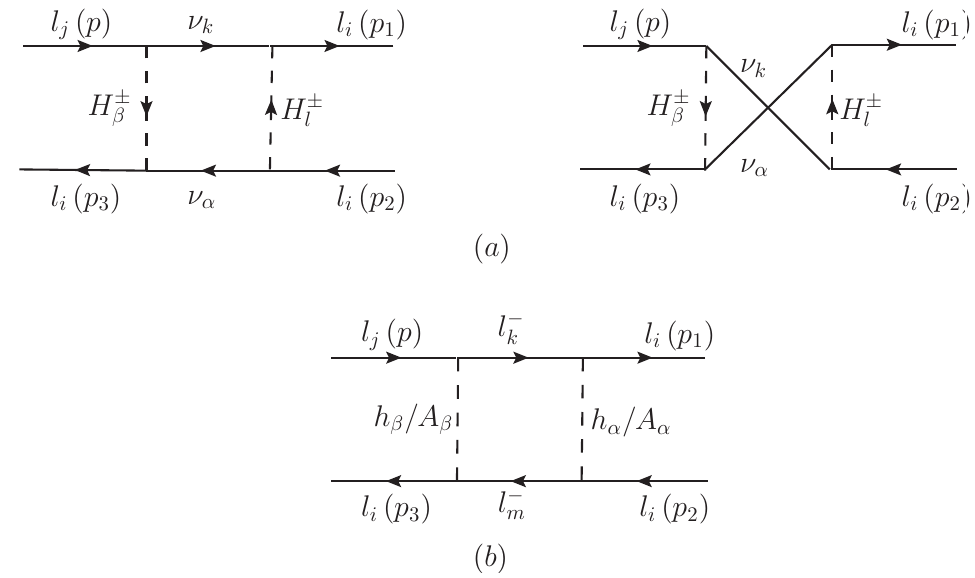}
\vspace{0cm}
\caption[]{Box-type diagrams contributing to the processes $l_j^-\rightarrow l_i^-l_i^-l_i^+$. (a) represents the contributions from neutral fermions $\nu_{k,\alpha}$ and charge scalars loops $H^\pm_{l,\beta}$, and (b) represents the contributions from charged fermions $l_{k,m}^-$ and neutral scalars $h_{\alpha,\beta}$ or pseudoscalars $A_{\alpha,\beta}$ loops.}
\label{figl3lbox}
\end{figure}

In addition, the box-type diagrams can also contribute to the processes $l_j^-\rightarrow l_i^-l_i^-l_i^+$. The corresponding Feynman diagrams are shown in Fig.~\ref{figl3lbox},
then the amplitude for the box-type diagrams can be written as
\begin{eqnarray}
&&T_{box}=\Big\{B_1^Le^2\bar u_i(p_1)\gamma_\mu P_Lu_j(p)\bar u_i(p_2)\gamma^\mu P_L\nu_i(p_3)+(L\leftrightarrow R)\Big\}\nonumber\\
&&\qquad\quad +\Big\{B_2^L[e^2\bar u_i(p_1)\gamma_\mu P_Lu_j(p)\bar u_i(p_2)\gamma^\mu P_R\nu_i(p_3)-(p_1\leftrightarrow p_2)]+(L\leftrightarrow R)\Big\}\nonumber\\
&&\qquad\quad +\Big\{B_3^L[e^2\bar u_i(p_1)P_Lu_j(p)\bar u_i(p_2)P_R\nu_i(p_3)-(p_1\leftrightarrow p_2)]+(L\leftrightarrow R)\Big\}\nonumber\\
&&\qquad\quad +\Big\{B_4^L[e^2\bar u_i(p_1)\sigma_{\mu,\nu}P_Lu_j(p)\bar u_i(p_2)\sigma^{\mu,\nu}P_L\nu_i(p_3)-(p_1\leftrightarrow p_2)]+(L\leftrightarrow R)\Big\},
\end{eqnarray}
where the coefficients $B_{1,2,3,4}^{L,R}$ originate from those box diagrams in Fig.~\ref{figl3lbox}, and the concrete expressions can be found in Appendix~\ref{wilsonl3l}. The decay width for $l_j^-\rightarrow l_i^-l_i^-l_i^+$ is~\cite{Hisano:1995cp}
\begin{eqnarray}
&&\Gamma(l_j^-\rightarrow l_i^-l_i^-l_i^+)=\frac{e^4m_{l_j}^5}{512\pi^3}\Big\{(|A_2^L|^2+|A_2^R|^2)\Big(\frac{16}{3}\ln\frac{m_{l_j}}{2m_{l_i}}-
\frac{14}{9}\Big)\nonumber\\
&&\qquad\quad\qquad\quad\qquad+(|A_1^L|^2+|A_1^R|^2)-2(A_1^LA_2^{R*}+A_2^LA_1^{R*}+H.c.)+\frac{1}{6}(|B_1^L|^2+|B_1^R|^2)
\nonumber\\
&&\qquad\quad\qquad\quad\qquad+\frac{1}{3}(|B_2^L|^2+|B_2^R|^2)+\frac{1}{24}(|B_3^L|^2+|B_3^R|^2)+6(|B_4^L|^2+|B_4^R|^2)
\nonumber\\
&&\qquad\quad\qquad\quad\qquad-\frac{1}{2}(B_3^LB_4^{L*}+B_3^RB_4^{R*}+H.c.)+\frac{1}{3}(A_1^LB_1^{L*}+A_1^RB_1^{R*}+
A_1^LB_2^{L*}\nonumber\\
&&\qquad\quad\qquad\quad\qquad+A_1^RB_2^{R*}+H.c.)-\frac{2}{3}(A_2^RB_1^{L*}+A_2^LB_1^{R*}+A_2^LB_2^{R*}+A_2^RB_2^{R*}+
H.c.)\nonumber\\
&&\qquad\quad\qquad\quad\qquad+\frac{1}{3}\Big[2(|F^{LL}|^2+|F^{RR}|^2)+(|F^{LR}|^2+|F^{RL}|^2)+(B_1^LF^{LL*}+
B_1^RF^{RR*}\nonumber\\
&&\qquad\quad\qquad\quad\qquad+B_2^LF^{LR*}+B_2^RF^{RL*}+H.c.)+2(A_1^LF^{LL*}+A_1^RLF^{RR*}+H.c.)\nonumber\\
&&\qquad\quad\qquad\quad\qquad+(A_1^LF^{LR*}+A_1^RLF^{RL*}+H.c.)-4(A_2^RF^{LL*}+A_2^LF^{RR*}+H.c.)\nonumber\\
&&\qquad\quad\qquad\quad\qquad-2(A_2^LF^{RL*}+A_2^RLF^{LR*}+H.c.)\Big]\Big\},
\end{eqnarray}
where
\begin{eqnarray}
&&F^{LL}=\sum_{N=Z,Z'}\frac{F^LC_{\bar l_i Nl_i}^L}{m_N^2},\;\;\;\;\;\;F^{RR}=F^{LL}(L\leftrightarrow R),\nonumber\\
&&F^{LR}=\sum_{N=Z,Z'}\frac{F^LC_{\bar l_i Nl_i}^R}{m_N^2},\;\;\;\;\;\;F^{RL}=F^{LR}(L\leftrightarrow R).
\end{eqnarray}

\section{Numerical analyses\label{sec4}}
\indent\indent

Considering the corrections at one loop level, we present the 
numerical results of muon anomalous MDM $\Delta a_\mu$ and the branching ratios of the LFV processes $l_j^-\rightarrow l_i^-\gamma$, $l_j^-\rightarrow l_i^-l_i^-l_i^+$ predicted in the FDM, in this section. The relevant SM input parameters are chosen as $m_W=80.385\;{\rm GeV},\;m_Z=90.1876\;{\rm GeV},\;\alpha_{em}(m_Z)=1/128.9$. The SM-like Higgs mass is~\cite{PDG}
\begin{eqnarray}
&&m_h=125.20\pm0.11\;{\rm GeV}.
\end{eqnarray} 
For the free parameters in the scalar sector of the FDM, we take $v_1=v_2$,  $\lambda_4'=\lambda_4''=\lambda_4/2$, $\lambda_5'=\lambda_5''=\lambda_5/2$, $\lambda_6'=\lambda_6''=\lambda_6/2$ and all parameters to be real for simplicity. 

The additional $Z'$ boson is subject to collider constraints. Recent ATLAS dilepton-resonance searches constrain its mass $M_{Z'}>4.05\;\text{TeV}$ at $95\%$ confidence level~\cite{ATLAS:2016cyf}, which is imposed in the numerical scan below. And the effects of $Z'$ depend directly on the relevant gauge coupling $g_F$, hence there is an lower bound on the ratio between $M_{Z'}$ and $g_F$, which is given at $99\%$ confidence level as~\cite{Cacciapaglia:2006pk,Carena:2004xs}
\begin{align}
    \frac{M_{Z'}}{g_F}>6\;\text{TeV}\label{MZpbound}.
\end{align}
For the charge normalization adopted here, we take $v_\chi\geq5\;{\rm TeV}$ in the numerical scan. Since $M_{Z'}\simeq2|z g_F|v_\chi$, this choice can well satisfy the lower bound on $M_{Z'}/g_F$ in Eq.~\eqref{MZpbound}.

\subsection{Muon MDM}

\begin{figure}[htbp]
	\setlength{\unitlength}{1mm}
	\centering
	\includegraphics[width=3.0in]{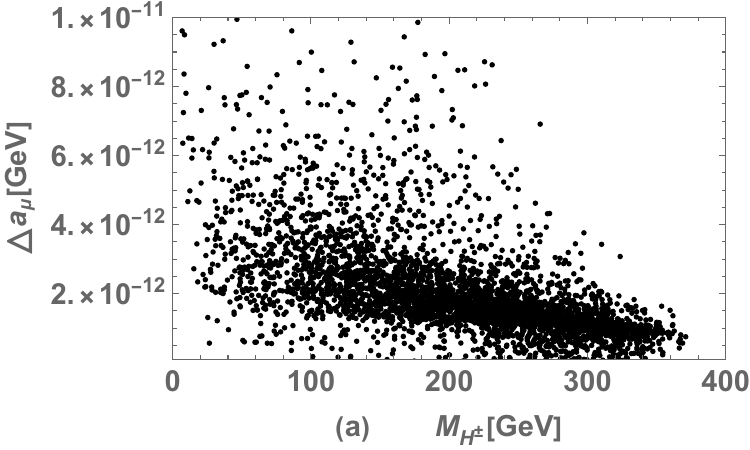}
	\vspace{0.5cm}
	\includegraphics[width=3.0in]{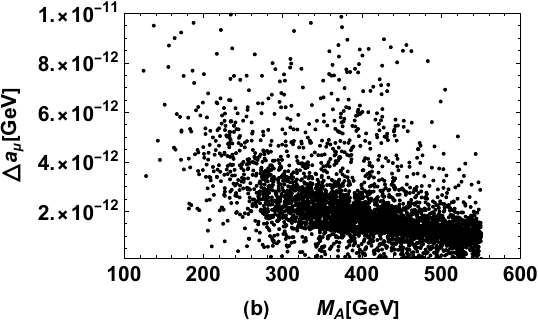}
	\vspace{0.5cm}
	\includegraphics[width=3.0in]{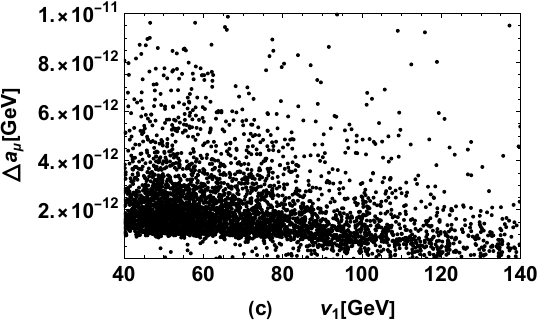}
	\vspace{0cm}
	\caption[]{The results of $\Delta a_\mu$  versus $M_{H^\pm}$, $M_{A}$, $v_1$ are
	plotted respectively by scanning the parameter space in Eq.~(\ref{eq38}). }
	\label{Camu}
\end{figure}

Considering the constraints from vacuum stability and perturbativity on the parameter space of FDM~\cite{Yang:2024znv,Cao:2025zwn}, we scan the parameter space
\begin{eqnarray}
	&&v_1=(0,\;140)\;{\rm GeV},\;\lambda_i=(0,\;5)\;\;{\rm with}\;\;(i=1,...,9,\chi),\;\lambda_{10}=(-4,\;0),\nonumber\\
	&&v_\chi=(5,\;40)\;{\rm TeV},\;\kappa=(-3,\;-0.1)\;{\rm TeV},\nonumber\\
	&&0<g_{_F}\leq0.6,\;
	g_{_{YF}}=(-0.8,\;0.8),\nonumber\\
	&&|m_{e,13}|=(0.0\sim0.3)\;{\rm GeV},\;\theta_{e,ij},\;(ij=12,\;13,\;23)=(-\pi\sim\pi)
	\label{eq38}
\end{eqnarray}
to compute the muon MDM numerically. The Yukawa structure is quite different in the model, which affects the theoretical predictions on Higgs signal strengths. Based on our analysis in Ref.~\cite{Cao:2025zwn}, the constraints from 125 GeV Higgs signal strengths in $2\sigma$~\cite{PDG}
\begin{eqnarray}
	&&\mu_{WW^*}=1.00\pm0.16,\;\mu_{ZZ^*}=1.02\pm0.16,\;\mu_{\gamma\gamma}=1.10\pm0.12,\;\mu_{b\bar b}=0.94\pm0.22,\;\nonumber\\
    &&\mu_{\mu^+\mu^-}=1.31\pm0.58,\;\mu_{\tau^+\tau^-}=0.91\pm0.18
\end{eqnarray}
and direct searches for additional scalars are imposed in the scan. Then the results of $\Delta a_\mu$ versus $M_{H^\pm}$, $\Delta a_\mu$ versus $M_{A}$, $\Delta a_\mu$ versus $v_1$ are plotted in Fig.~\ref{Camu} (a), (b), (c) respectively.

Fig.~\ref{Camu} clearly demonstrates that the new contributions in the FDM can well satisfy the latest experimental result on $\Delta a_\mu$, which exhibits a monotonically decreasing trend with increasing $M_{H^\pm}$ as shown in Fig.~\ref{Camu} (a). And for relatively light $H^\pm$ ($0\;{\rm GeV}<M_{H^\pm}<100\;{\rm GeV}$), the values of $\Delta a_\mu$ exhibit a wider dispersion. Fig.~\ref{Camu} (b) and (c) indicate the effects of $M_A$, $v_1$ are similar to the ones of $M_{H^\pm}$, but weaker than the ones of $M_{H^\pm}$. This fact results from that the dominant contributions to $\Delta a_\mu$ in the FDM come from the $H^\pm$-mediated loop diagram. Hence the contributions are suppressed by heavy $H^\pm$.

\subsection{Branching ratios for LFV processes}

\begin{figure}
\setlength{\unitlength}{1mm}
\centering
\includegraphics[width=3.1in]{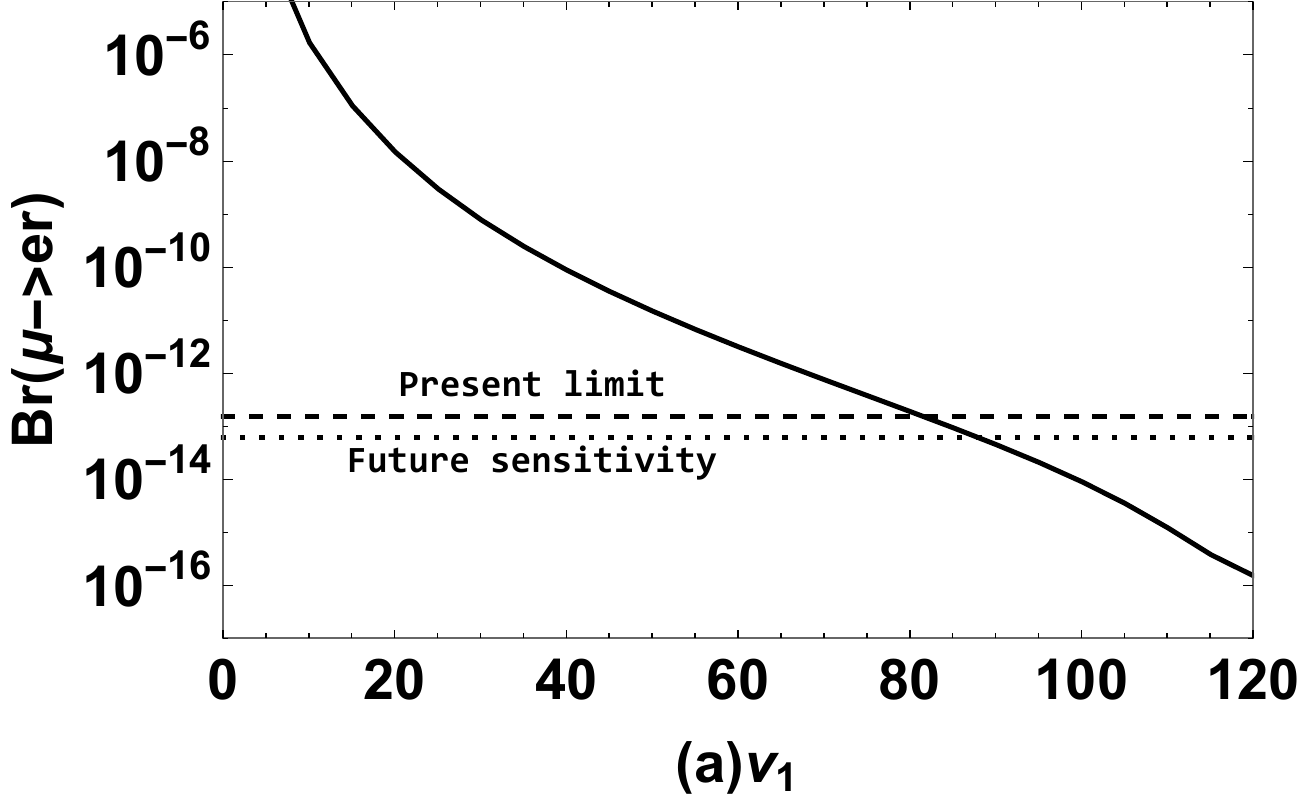}%
\vspace{0.5cm}
\includegraphics[width=3.1in]{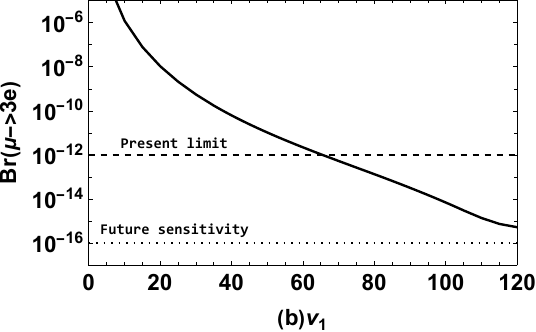}
\vspace{0cm}
\par
\hspace{-0.in}
\includegraphics[width=3.1in]{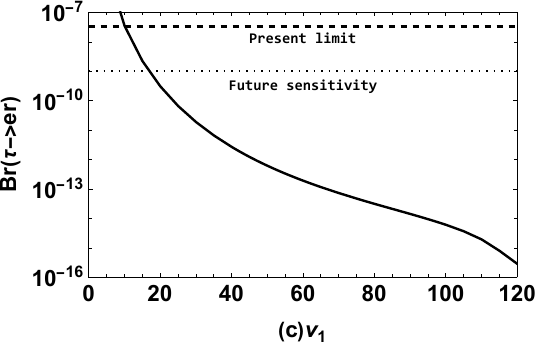}%
\vspace{0.5cm}
\includegraphics[width=3.1in]{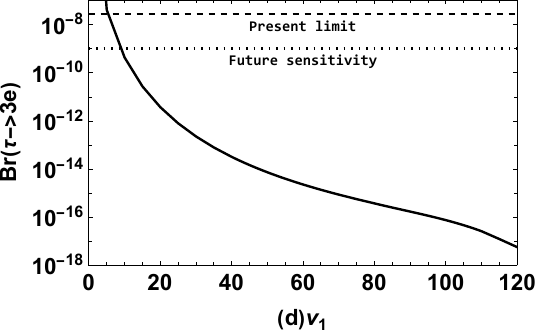}
\vspace{0cm}
\par
\hspace{-0.in}
\includegraphics[width=3.1in]{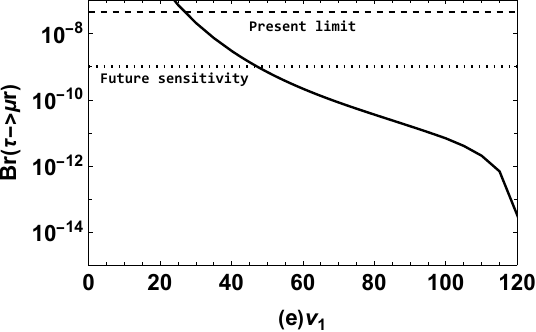}%
\vspace{0.5cm}
\includegraphics[width=3.1in]{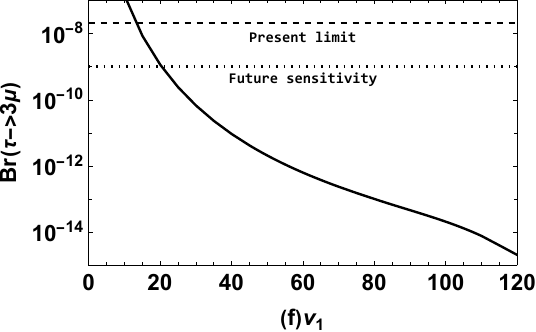}
\vspace{0cm}
\caption[]{LFV rates for $l_j-l_i$ transitions versus $v_1$ are plotted, where the dashed and dotted lines denote the present limits and future sensitivities respectively.}
\label{Cljli}
\end{figure}
Finally, we present the numerical results of the branching ratios for the processes $l_{j}^{-}\to l_{i}^{-}\gamma $ and $l_{j}^{-}\to l_{i}^{-}l_{i}^{-}l_{i}^{+}$. In order to carry out the numerical computations, we take the following parameter space
\begin{eqnarray}
	&&\lambda_1=4.80,\;\lambda_2=3.50,\; \lambda_3=2.70,\;\lambda_4=2.50,\;\lambda_5=0.30,\;\lambda_6=3.00,\;\lambda_7=4.90,\nonumber\\
    &&\lambda_8=3.00,\;\lambda_9=2.70,\;\lambda_{10}=-2.00,\;\lambda_\chi=2.50,\;v_\chi=8.70\;{\rm TeV},\;\kappa=-0.20\;{\rm TeV},\nonumber\\
    &&g_{_F}=0.40,\;g_{_{YF}}=0.20,
\end{eqnarray}
which yields the observed $125$ GeV Higgs mass and the latest experimental measurements of the muon $g-2$. The fixed parameters listed above have negligible effects on the LFV results. Since the Yukawa coupling matrix ${{Y}_{e}}$ has a significant impact on the branching ratios of LFV processes, we have investigated the variation of the branching ratios of LFV processes with respect to the elements of the ${{Y}_{e}}$ matrix, which can be derived through Eq.~(\ref{eqme}). The VEV of the Higgs doublet and the matrix elements of the electron mass matrix both influence the Yukawa couplings, the influence of $\theta_{e,ij}$ on the electron mass matrix is given in Eq.~(\ref{eq13}). Therefore, we respectively choose $v_1,\;m_{e,13},\;\theta_{e,ij}$ as variables for investigation.

We choose $m_{e,13}=0.04,\;\theta_{e,12}=0.40\pi,\;\theta_{e,13}=-0.80\pi,\;\theta_{e,23}=-0.45\pi$. Then we plot LFV rates for $l_j-l_i$ transitions versus $v_1$ in Fig.~\ref{Cljli}. The dashed and dotted lines denote the present limits and future sensitivities respectively. It is obvious that the LFV rates decrease with the increasing of $v_1$, which indicates that a larger $v_1$ suppresses the LFV rates. Fig.~\ref{Cljli} (a, b) shows that $\text{Br}(\mu\rightarrow e\gamma)$ and  $\text{Br}(\mu\rightarrow 3e)$ are above the corresponding experimental upper bounds for small $v_1$. And the future experiments have great opportunity to observe these two processes. In addition, Fig.~\ref{Cljli} (c-f) show a similar trend for the predictions of the branching ratios of the other four processes, but the constraints are weaker than the ones from $\text{Br}(\mu\rightarrow e\gamma)$ and  $\text{Br}(\mu\rightarrow 3e)$.

\begin{figure}[!h]
\setlength{\unitlength}{1mm}
\centering
\includegraphics[width=3.1in]{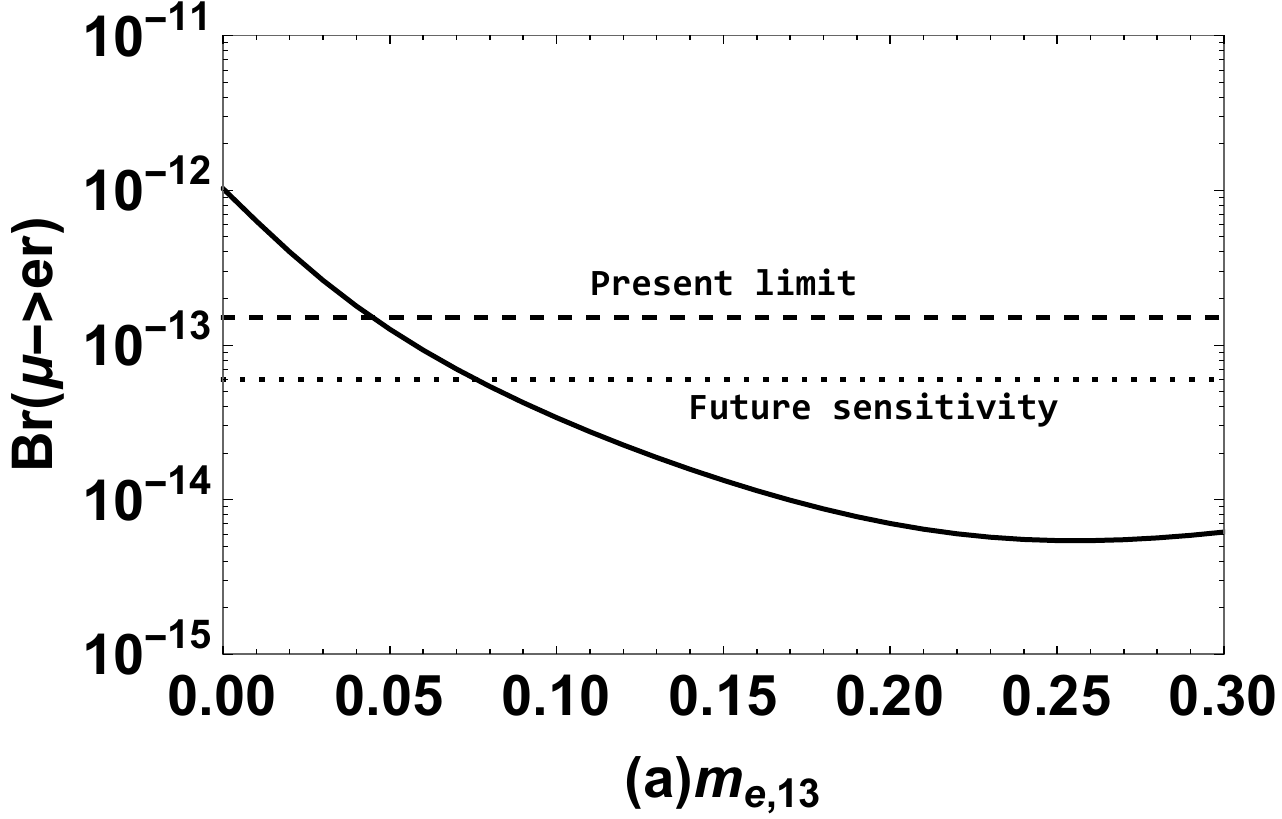}%
\vspace{0.5cm}
\includegraphics[width=3.1in]{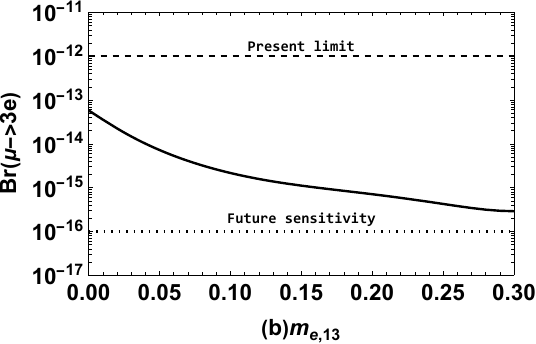}
\vspace{0cm}
\par
\hspace{-0.in}
\includegraphics[width=3.1in]{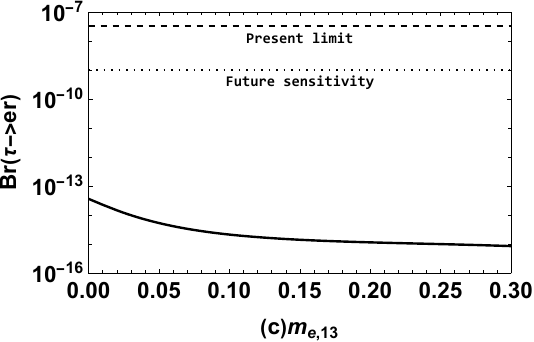}%
\vspace{0.5cm}
\includegraphics[width=3.1in]{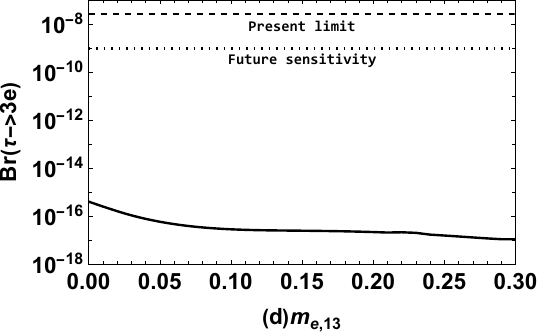}
\vspace{0cm}
\par
\hspace{-0.in}
\includegraphics[width=3.1in]{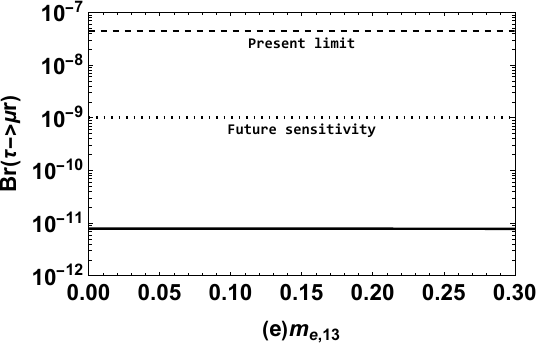}%
\vspace{0.5cm}
\includegraphics[width=3.1in]{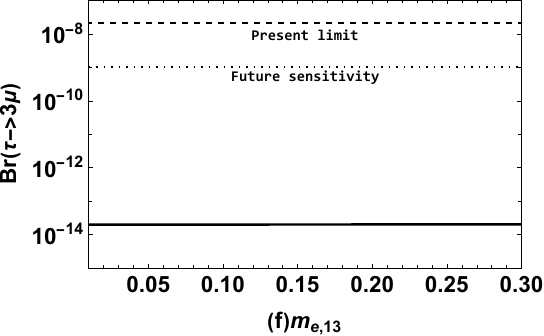}
\vspace{0cm}
\caption[]{Branching ratios for $l_j-l_i$ versus $m_{e,13}$ are plotted, where the dashed and dotted lines denote the present limits and future sensitivities respectively.}
\label{CljliME}
\end{figure}
Then we appropriately fix $v_1=120{\rm GeV},\;\theta_{e,ij}$ to explore the effects of $m_{e,13}$ on the branching ratios for LFV transitions, and $\theta_{e,ij}$ is also taken as above. The numerical results versus $m_{e,13}$ are in Fig.~\ref{CljliME}. Similarly, the dashed and dotted lines denote the present limits and future sensitivities respectively. From Fig.~\ref{CljliME} (a, b), it is evident that the parameter $m_{e,13}$ exhibits a strong correlation with the branching ratio of $\text{Br}(\mu\rightarrow e\gamma)$ and $\text{Br}(\mu\rightarrow 3e)$. Specifically, as $m_{e,13}$ increases, the branching ratios for these processes decrease, thereby exerting a significant suppressive effect on the overall rates of LFV phenomena. However, this feature does not appear in Fig.~\ref{CljliME} (c-f). Fig.~\ref{CljliME} (c, d) show that as the parameter $m_{e,13}$ increases, the branching ratios $\text{Br}(\tau\rightarrow e\gamma)$ and $\text{Br}(\tau\rightarrow 3e)$ decrease slowly, while $\text{Br}(\tau\rightarrow mu\gamma)$ and $\text{Br}(\tau\rightarrow 3\mu)$ depend only weakly on $m_{e,13}$.

\begin{figure}[!h]
	\setlength{\unitlength}{1mm}
	\centering
	\includegraphics[width=3.1in]{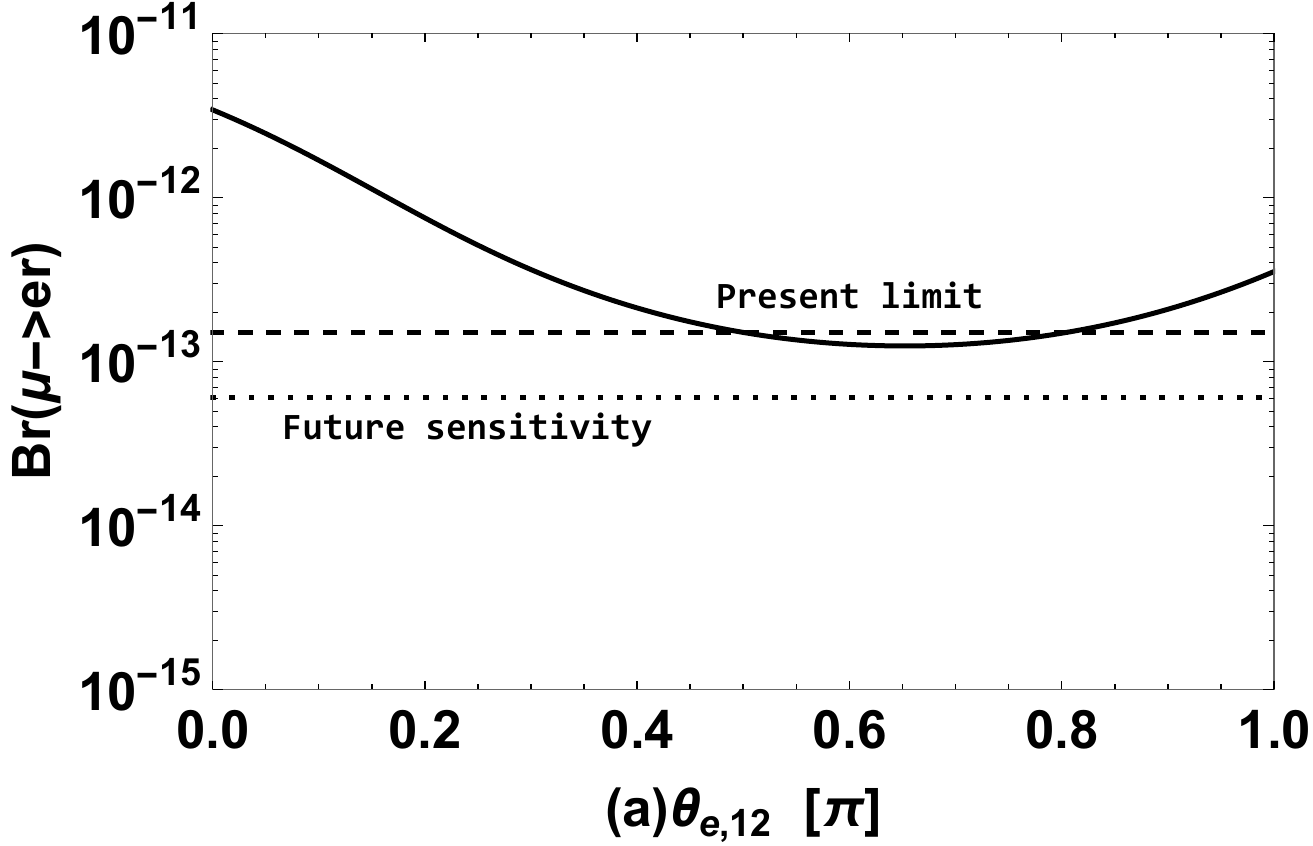}%
	\vspace{0.5cm}
	\includegraphics[width=3.1in]{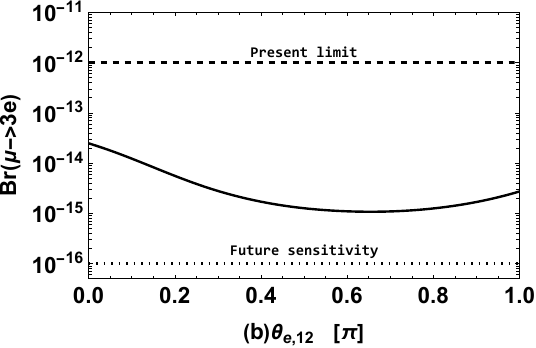}
	\vspace{0cm}
	\par
	\hspace{-0.in}
	\includegraphics[width=3.1in]{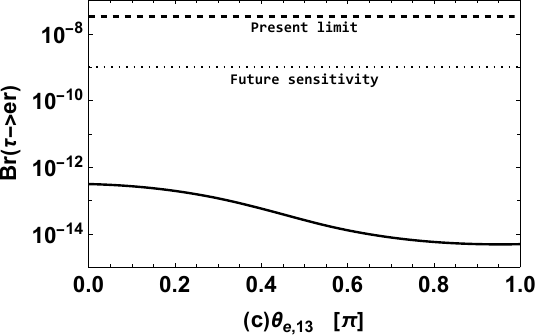}%
	\vspace{0.5cm}
	\includegraphics[width=3.1in]{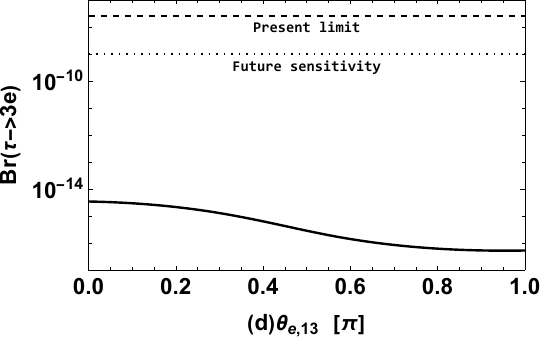}
	\vspace{0cm}
	\par
	\hspace{-0.in}
	\includegraphics[width=3.1in]{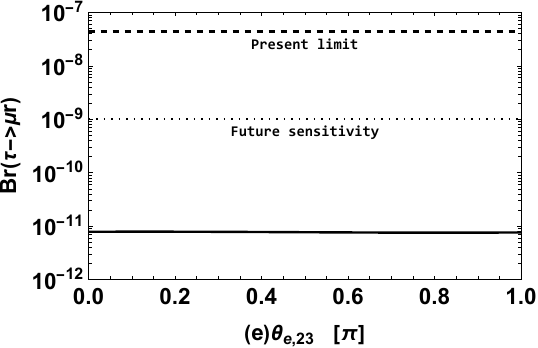}%
	\vspace{0.5cm}
	\includegraphics[width=3.1in]{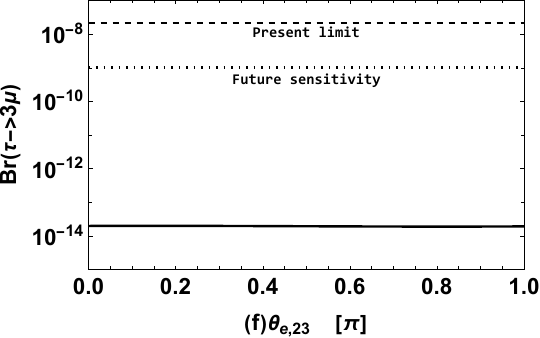}
	\vspace{0cm}
	\caption[]{LFV rates for $l_j-l_i$ transitions versus $\theta_{e,ij}$ are plotted, where the dashed and dotted lines denote the present limits and future sensitivities respectively.}
	\label{CljliIME}
\end{figure}

In order to see the effects of $\theta_{e,ij}$, we appropriately fix $v_1=120\;{\rm GeV}$ and $m_{e,13}=0.040$, then we plot $\text{Br}(\mu\rightarrow e\gamma)$ and $\text{Br}(\mu\rightarrow 3e)$ versus $\theta_{e,12}$ for $\theta_{e,13}=-0.80\pi,\;\theta_{e,23}=-0.45\pi$ in Fig.~\ref{CljliIME} (a, b). In Fig.~\ref{CljliIME} (c, d) we plot $\text{Br}(\tau\rightarrow e\gamma)$ and $\text{Br}(\tau\rightarrow 3e)$ versus $\theta_{e,13}$ for $\theta_{e,12}=0.40\pi,\;\theta_{e,23}=-0.45\pi$. $\text{Br}(\tau\rightarrow \mu\gamma)$ and $\text{Br}(\tau\rightarrow 3\mu)$ versus $\theta_{e,23}$ for $\theta_{e,12}=0.40\pi,\;\theta_{e,13}=-0.80\pi$ are shown in Fig.~\ref{CljliIME} (e, f). For each figure, the fixed phases affect the corresponding numerical results negligibly. The plots show that the CPV phases can affect the theoretical predictions on $\text{Br}(\mu\rightarrow e\gamma)$, $\text{Br}(\mu\rightarrow 3e)$, $\text{Br}(\tau\rightarrow e\gamma)$ and $\text{Br}(\tau\rightarrow 3e)$ by influencing the mixing strength between different flavors, while they affect $\text{Br}(\tau\rightarrow \mu\gamma)$ and $\text{Br}(\tau\rightarrow 3\mu)$ negligibly.

\section{Summary\label{sec5}}
\indent\indent
Motivated by the new flavor changing neutral couplings in the flavor-dependent $U(1)_F$ model, we investigate the predicted muon anomalous magnetic moment and charged lepton flavor violation decays. The numerical results indicate that most of the model parameter space can well satisfy the latest experimental data on the muon anomalous magnetic moment. Then considering the obtained constraints, we present the predicted branching ratios of charged lepton flavor changing (LFV) decays. The experimental upper bounds on these LFV processes set strict limits on $v_1$, and $|m_{e,13}|$, $\theta_{e,ij}$ also influence the theoretical predictions for $\text{Br}(\mu\rightarrow e\gamma)$, $\text{Br}(\mu\rightarrow 3e)$, $\text{Br}(\tau\rightarrow e\gamma)$ and $\text{Br}(\tau\rightarrow 3e)$. In addition, the numerical results imply that these LFV processes have a good prospect to be observed at the next generation experiments.

\begin{acknowledgments}
\indent\indent
The work has been supported by the National Natural Science Foundation of China (NNSFC) with Grants No. 11535002, No. 11647120, and No. 11705045, Natural Science Foundation of Hebei province with Grants No. A2016201010 and No. A2016201069, Hebei Key Lab of Optic-Eletronic Information and Materials, and the Midwest Universities Comprehensive Strength Promotion
project.
\end{acknowledgments}

\appendix

\section{The Wilson coefficients of the process $l_j^-\rightarrow l_i^-\gamma$. \label{wilsonllr}}
The coefficients corresponding to Fig.~\ref{figllr}(a), (b) can be written as
\begin{eqnarray}
	&&A_1^{(a)L}=\frac{1}{6m_W^2}C_{\bar l_i \nu_{k} H^\pm_{l} }^L C_{\bar \nu_{k} H^{\pm*}_{l} l_j}^R I_4(x_{_{\nu_{k}}},x_{_{H^\pm_{l}}}),\nonumber\\
	&&A_2^{(a)L}=\frac{m_{\nu_{k}}}{m_{l_j}m_W^2}C_{\bar l_i \nu_{k} H^\pm_{l}}^L C_{\bar \nu_{k} H^{\pm*}_{l} l_j}^L [I_3(x_{_{\nu_{k}}},x_{_{H^\pm_{l}}})-I_1(x_{_{\nu_{k}}},x_{_{H^\pm_{l}}})],\nonumber\\
	&&A_{1,2}^{(a)R}=A_{1,2}^{(a)L}(L\leftrightarrow R),\nonumber\\
	&&A_1^{(b)L}=\frac{1}{6m_W^2}C_{\bar l_i l_k^- h_{\alpha}}^R C_{\bar l_k^- h_{\alpha} l_j}^L [I_1(x_{_{l_k^-}},x_{_{h_{\alpha}}})-2I_2(x_{_{l_k^-}},x_{_{h_{\alpha}}})-I_4(x_{_{l_k^-}},x_{_{h_{\alpha}}})],({h_{\alpha}}\leftrightarrow {A_{\alpha}})\nonumber\\
	&&A_2^{(b)L}=\frac{m_{l_k^-}}{m_{l_j}m_W^2}C_{\bar l_i l_k^- h_{\alpha}}^L C_{\bar l_k^- h_{\alpha} l_j}^L [I_1(x_{_{l_k^-}},x_{_{h_{\alpha}}})-I_2(x_{_{l_k^-}},x_{_{h_{\alpha}}})-I_4(x_{_{l_k^-}},x_{_{h_{\alpha}}})],({h_{\alpha}}\leftrightarrow {A_{\alpha}})\nonumber\\
	&&A_{1,2}^{(b)R}=A_{1,2}^{(b)L}(L\leftrightarrow R).
\end{eqnarray}
where $x_i=m_i^2/m_W^2$, and the concrete expressions for the functions $I_{1,2,3,4}$ and $G_{1,2,3,4}$ below can be found in Ref.~\cite{Zhang:2013hva,Zhang:2013jva}.

The related coupling vertexes are
\begin{align}
C_{\bar l_i \nu_{k} H^\pm_{l}}^L
&=i\biggl(U_{L,j3}^{e,*}\Bigl(Y_{\nu31}^{D}U_{i4}^{V,*}Z_{k2}^{+}+Y_{\nu32}^{D}U_{i5}^{V,*}Z_{k1}^{+}\Bigr)
+\Bigl(Y_{\nu12}^{D}U_{L,j1}^{e,*}U_{i5}^{V,*}+Y_{\nu21}^{D}U_{L,j2}^{e,*}U_{i4}^{V,*}\Bigr)Z_{k3}^{+}\biggr)\nonumber\\
C_{\bar l_i \nu_{k} H^\pm_{l}}^R
&=i\biggl(-Y_{e13}^{*}U_{R,j3}^{e}U_{i1}^{V}Z_{k1}^{+}
-Y_{e31}^{*}U_{R,j1}^{e}U_{i3}^{V}Z_{k1}^{+}
-Y_{e23}^{*}U_{R,j3}^{e}U_{i2}^{V}Z_{k2}^{+}
-Y_{e32}^{*}U_{R,j2}^{e}U_{i3}^{V}Z_{k2}^{+}\nonumber\\
&\quad -Y_{e12}^{*}U_{R,j2}^{e}U_{i1}^{V}Z_{k3}^{+}
-Y_{e21}^{*}U_{R,j1}^{e}U_{i2}^{V}Z_{k3}^{+}
-Y_{e33}^{*}U_{R,j3}^{e}U_{i3}^{V}Z_{k3}^{+}\biggr)
\end{align}

\begin{align}
C_{\bar \nu_{k} H^{\pm*}_{l} l_j}^L
&=i\biggl(-U_{R,i2}^{e,*}\Bigl(Y_{e12}U_{j1}^{V,*}Z_{k3}^{+}+Y_{e32}U_{j3}^{V,*}Z_{k2}^{+}\Bigr)
-U_{R,i1}^{e,*}\Bigl(Y_{e21}U_{j2}^{V,*}Z_{k3}^{+}+Y_{e31}U_{j3}^{V,*}Z_{k1}^{+}\Bigr)\nonumber\\
&-U_{R,i3}^{e,*}\Bigl(Y_{e13}U_{j1}^{V,*}Z_{k1}^{+}+Y_{e23}U_{j2}^{V,*}Z_{k2}^{+}+Y_{e33}U_{j3}^{V,*}Z_{k3}^{+}\Bigr)\biggr)\nonumber\\
C_{\bar \nu_{k} H^{\pm*}_{l} l_j}^R
&=i\biggl(\Bigl(Y_{\nu12}^{D,*}U_{L,i1}^{e}U_{j5}^{V}+Y_{\nu21}^{D,*}U_{L,i2}^{e}U_{j4}^{V}\Bigr)Z_{k3}^{+}
+Y_{\nu31}^{D,*}U_{L,i3}^{e}U_{j4}^{V}Z_{k2}^{+}
+Y_{\nu32}^{D,*}U_{L,i3}^{e}U_{j5}^{V}Z_{k1}^{+}\biggr)
\end{align}

\begin{align}
C_{\bar l_i l_k^- h_{\alpha}}^L
&=\frac{-i}{\sqrt{2}}\biggl(U_{R,i2}^{e,*}\Bigl(Y_{e12}U_{L,j1}^{e,*}Z_{k3}^{H}+Y_{e32}U_{L,j3}^{e,*}Z_{k2}^{H}\Bigr)
+U_{R,i1}^{e,*}\Bigl(Y_{e21}U_{L,j2}^{e,*}Z_{k3}^{H}+Y_{e31}U_{L,j3}^{e,*}Z_{k1}^{H}\Bigr)\nonumber\\
&+U_{R,i3}^{e,*}\Bigl(Y_{e13}U_{L,j1}^{e,*}Z_{k1}^{H}+Y_{e23}U_{L,j2}^{e,*}Z_{k2}^{H}+Y_{e33}U_{L,j3}^{e,*}Z_{k3}^{H}\Bigr)\biggr)\nonumber\\
C_{\bar l_i l_k^- h_{\alpha}}^R
&=\frac{-i}{\sqrt{2}}\biggl(Y_{e13}^{*}U_{R,j3}^{e}U_{L,i1}^{e}Z_{k1}^{H}
+Y_{e31}^{*}U_{R,j1}^{e}U_{L,i3}^{e}Z_{k1}^{H}
+Y_{e23}^{*}U_{R,j3}^{e}U_{L,i2}^{e}Z_{k2}^{H}
+Y_{e32}^{*}U_{R,j2}^{e}U_{L,i3}^{e}Z_{k2}^{H}\nonumber\\
&\qquad +Y_{e12}^{*}U_{R,j2}^{e}U_{L,i1}^{e}Z_{k3}^{H}
+Y_{e21}^{*}U_{R,j1}^{e}U_{L,i2}^{e}Z_{k3}^{H}
+Y_{e33}^{*}U_{R,j3}^{e}U_{L,i3}^{e}Z_{k3}^{H}\biggr)
\end{align}

\begin{align}
C_{\bar l_i l_k^- A_{\alpha}}^L
&=-\frac{1}{\sqrt{2}}\biggl(U_{R,i2}^{e,*}\Bigl(Y_{e12}U_{L,j1}^{e,*}Z_{k3}^{A}+Y_{e32}U_{L,j3}^{e,*}Z_{k2}^{A}\Bigr)
+U_{R,i1}^{e,*}\Bigl(Y_{e21}U_{L,j2}^{e,*}Z_{k3}^{A}+Y_{e31}U_{L,j3}^{e,*}Z_{k1}^{A}\Bigr)\nonumber\\
&\qquad +U_{R,i3}^{e,*}\Bigl(Y_{e13}U_{L,j1}^{e,*}Z_{k1}^{A}+Y_{e23}U_{L,j2}^{e,*}Z_{k2}^{A}+Y_{e33}U_{L,j3}^{e,*}Z_{k3}^{A}\Bigr)\biggr)\nonumber\\
C_{\bar l_i l_k^- A_{\alpha}}^R
&=\frac{1}{\sqrt{2}}\biggl(Y_{e13}^{*}U_{R,j3}^{e}U_{L,i1}^{e}Z_{k1}^{A}
+Y_{e31}^{*}U_{R,j1}^{e}U_{L,i3}^{e}Z_{k1}^{A}
+Y_{e23}^{*}U_{R,j3}^{e}U_{L,i2}^{e}Z_{k2}^{A}
+Y_{e32}^{*}U_{R,j2}^{e}U_{L,i3}^{e}Z_{k2}^{A}\nonumber\\
&\qquad +Y_{e12}^{*}U_{R,j2}^{e}U_{L,i1}^{e}Z_{k3}^{A}
+Y_{e21}^{*}U_{R,j1}^{e}U_{L,i2}^{e}Z_{k3}^{A}
+Y_{e33}^{*}U_{R,j3}^{e}U_{L,i3}^{e}Z_{k3}^{A}\biggr)
\end{align}

\section{The loop functions $I_N$ and $J_N$ . \label{au}}
In this appendix, we show the used formulae for the functions $I_N$
and $J_N$, which are defined as
\begin{eqnarray}
I_N(m_1^2,\cdots,m_N^2) &=&
\int\frac{d^4k}{(2\pi)^4i}
\frac{1}{(k^2-m_1^2)\cdots (k^2-m_N^2)},
\label{ap_I_N} \\
J_N(m_1^2,\cdots,m_N^2) &=&
\int\frac{d^4k}{(2\pi)^4i}
\frac{k^2}{(k^2-m_1^2)\cdots (k^2-m_N^2)}.
\label{ap_J_N}
\end{eqnarray}
The signs of the functions $I_N$ and $J_N$ are given by
\begin{eqnarray}
(-1)^{N} I_N(m_1^2,\cdots,m_N^2) &>& 0,
\\
(-1)^{N+1} J_N(m_1^2,\cdots,m_N^2) &>& 0,
\end{eqnarray}
The functions $I_N$ and $I_{N-1}$ are related as
\begin{eqnarray}
I_N(m_1^2,\cdots,m_N^2) = \frac{1}{m_1^2-m_N^2}
\{ I_{N-1}(m_1^2,\cdots,m_{N-1}^2) - I_{N-1}(m_2^2,\cdots,m_N^2) \},
\label{IN>=3}
\end{eqnarray}
and the explicit form of $I_2$ is given by
\begin{eqnarray}
I_2(m_1^2,m_2^2) = -\frac{1}{16\pi^2}
\left\{ \frac{m_1^2}{m_1^2-m_2^2} \ln\left( \frac{m_1^2}{\Lambda^2} \right)
+ \frac{m_2^2}{m_2^2-m_1^2} \ln\left( \frac{m_2^2}{\Lambda^2} \right)
\right\}.
\label{I_2}
\end{eqnarray}
Notice that the function $I_2$ is logarithmically divergent, and hence
$I_2$ defined in Eq.~(\ref{I_2}) depends on a cut-off parameter
$\Lambda$. However, $I_N$ ($N\geq 3$) iteratively defined by
using Eq.~(\ref{IN>=3}) is independent of $\Lambda$, as it should be.
In addition, $J_N$ is related to $I_N$ and $I_{N-1}$ as
\begin{eqnarray}
J_N(m_1^2,\cdots,m_N^2) = I_{N-1}(m_1^2,\cdots,m_{N-1}^2)
+ m_N^2 I_N(m_1^2,\cdots,m_N^2).
\end{eqnarray}
In the case where all the masses $m_1$ -- $m_N$ are almost degenerate,
it is convenient to use the Taylar expansion of $I_N$. Define
\begin{eqnarray}
\epsilon_i \equiv \frac{\bar{m}^2-m_i^2}{\bar{m}^2}~~~(i=1-N),
\end{eqnarray}
with $\bar{m}$ being an arbitrary mass scale, then $I_N$ is expanded as
\begin{eqnarray}
I_N(m_1^2,\cdots,m_N^2) &=& \frac{(-1)^N}{16\pi^2}
\frac{1}{\bar{m}^{2(N-2)}} \sum_{p=0}^\infty
\frac{1}{(N+p-2)(N+p-1)}
\nonumber \\ &&
\times \sum_{j_1+\cdots +j_N=p}
\epsilon_1^{j_1}\cdots\epsilon_N^{j_N}~~~(N\geq 3),
\end{eqnarray}
and for $N=2$,
\begin{eqnarray}
I_2(m_1^2,m_2^2) = - \frac{1}{16\pi^2}
\left\{ \ln\left( \frac{\bar{m}^2}{\Lambda^2}\right) +1 \right\}
+ \frac{1}{16\pi^2}
\sum_{p=1}^\infty
\frac{1}{p(p+1)}
\sum_{j_1+j_2=p}
\epsilon_1^{j_1}\epsilon_2^{j_2}.
\label{Taylar_I2}
\end{eqnarray}
Notice that Eqs.~(\ref{IN>=3}) -- (\ref{Taylar_I2}) are used in the numerical calculations.

Furthermore, the function $I_N$ has mass dimension $(4-2N)$. Therefore, we obtain
\begin{eqnarray}
\frac{d}{d\lambda}\left\{
\lambda^{2-N} I_N(\lambda m_1^2,\cdots,\lambda m_N^2)  \right\}
= 0,
\end{eqnarray}
which reduces to
\begin{eqnarray}
(2-N) I_N(m_1^2,\cdots,m_N^2) +
\sum_{i=1}^{N} m_i^2
I_{N+1} (m_1^2,\cdots,m_i^2,m_i^2,\cdots,m_N^2) =0.
\end{eqnarray}
Similar formula can be obtained for $J_N$;
\begin{eqnarray}
(3-N) J_N(m_1^2,\cdots,m_N^2) +
\sum_{i=1}^{N} m_i^2
J_{N+1} (m_1^2,\cdots,m_i^2,m_i^2,\cdots,m_N^2) =0.	
\end{eqnarray}

\section{The Wilson coefficients of the process $l_j^-\rightarrow l_i^-l_i^-l_i^+$. \label{wilsonl3l}}
The coefficients corresponding to N-penguin contributions can be written as
\begin{eqnarray}
	&&F^L=\frac{1}{2e^2}C_{\bar l_i \nu_{k} H^\pm_{l}}^RC_{H^{\pm*}_{l}NH^\pm_{\beta }}^RC_{\bar \nu_{k} H^{\pm*}_{\beta} l_j}G_2(x_{\nu_{k}},
	x_{H^\pm_{\beta}},x_{H^\pm_{l}})\nonumber\\
	&&\qquad\quad+\frac{m_{l_k^-}m_{l_m^-}}{e^2m_W^2}C_{\bar l_i h_{\beta} l_k^-}^RC_{\bar l_k^- N l_m^-}^L
	C_{\bar l_m^- h_{\beta} l_j}^LG_1(x_{h_{\beta}},x_{l_k^-},x_{l_m^-})\nonumber\\
	&&\qquad\quad-\frac{1}{2e^2}C_{\bar l_i h_{\beta } l_k^-}^RC_{\bar l_k^- N l_m^-}^RC_{\bar l_m^- h_{\beta} l_j}^LG_2(x_{h_{\beta}},x_{l_k^-},x_{l_m^-}),\nonumber\\
	&&F^R=F^L({L\leftrightarrow R}),
\end{eqnarray}

The coefficients corresponding to box-type diagrams are
\begin{eqnarray}
	&&B_1^L=\frac{m_{\nu_{k}}m_{\nu_{\alpha}}}{e^2m_W^2}G_3(x_{\nu_{k}},x_{\nu_{\alpha}},x_{H^\pm_{\beta}},x_{H^\pm_{l}})C_{\bar l_i H^\pm_{l} \nu_{k}}^LC_{\bar \nu_{k} H^{\pm*}_{\beta} l_j}^LC_{\bar l_i H^\pm_{l} \nu_{\alpha}}^RC_{\bar \nu_{\alpha} H^{\pm*}_{\beta} l_i}^R\nonumber\\
	&&\qquad\quad+\frac{1}{2e^2m_W^2}G_4(x_{\nu_{k}},x_{\nu_{\alpha}},x_{H^\pm_{\beta}},x_{H^\pm_{l}})[C_{\bar l_i H^\pm_{l} \nu_{k}}^RC_{\bar \nu_{k} H^{\pm*}_{\beta} l_j}^LC_{\bar l_i H^\pm_{\beta} \nu_{\alpha}}^RC_{\bar \nu_{\alpha} H^{\pm*}_{l} l_i}^L\nonumber\\
	&&\qquad\quad+C_{\bar l_i H^\pm_{l} \nu_{k}}^LC_{\bar \nu_{k} H^{\pm*}_{\beta} l_j}^RC_{\bar l_i H^\pm_{\beta} \nu_{\alpha}}^RC_{\bar \nu_{\alpha} H^{\pm*}_{l}l_i}^L]+\frac{1}{2e^2m_W^2}G_4(x_{l_k^-},x_{l_m^-},x_{h_{\beta}},x_{h_{\alpha}})\nonumber\\
	&&\qquad\quad\times C_{\bar l_i h_{\alpha} l_k^-}^RC_{\bar l_k^- h_{\beta} l_j}^LC_{\bar l_i h_{\beta} l_m^-}^RC_{\bar l_m^- h_{\alpha} l_i}^L,\nonumber\\
	&&B_2^L=-\frac{m_{\nu_{k}}m_{\nu_{\alpha}}}{2e^2m_W^2}G_3(x_{\nu_{k}},x_{\nu_{\alpha}},x_{H^\pm_{\beta}},x_{H^\pm_{l}})C_{\bar l_i H^\pm_{l} \nu_{k}}^RC_{\bar \nu_{k} H^{\pm*}_{\beta} l_j}^RC_{\bar l_i H^\pm_{\beta} \nu_{\alpha}}^LC_{\bar \nu_{\alpha} H^{\pm*}_{l} l_i}^L\nonumber\\
	&&\qquad\quad+\frac{1}{4e^2m_W^2}G_4(x_{\nu_{k}},x_{\nu_{\alpha}},x_{H^\pm_{\beta}},x_{H^\pm_{l}})[C_{\bar l_i H^\pm_{l} \nu_{k}}^RC_{\bar \nu_{k} H^{\pm*}_{\beta} l_j}^LC_{\bar l_i H^\pm_{\beta} \nu_{\alpha}}^LC_{\bar \nu_{\alpha} H^{\pm*}_{l} l_i}^R\nonumber\\
	&&\qquad\quad+C_{\bar l_i H^\pm_{l} \nu_{k}}^RC_{\bar \nu_{k} H^{\pm*}_{\beta} l_j}^LC_{\bar l_i H^\pm_{l} \nu_{\alpha}}^RC_{\bar \nu_{\alpha} H^{\pm*}_{\beta}l_i}^L]+\frac{1}{4e^2m_W^2}G_4(x_{l_k^-},x_{l_m^-},x_{h_{\beta}},x_{h_{\alpha}})\nonumber\\
	&&\qquad\quad\times C_{\bar l_i h_{\alpha} l_k^-}^RC_{\bar l_k^- h_{\beta} l_j}^LC_{\bar l_i h_{\beta} l_m^-}^LC_{\bar l_m^- h_{\alpha} l_i}^R-\frac{m_{l_k^-}m_{l_m^-}}{2e^2m_W^2}G_3(x_{l_k^-},x_{l_m^-},x_{h_{\beta}},x_{h_{\alpha}})\nonumber\\
	&&\qquad\quad\times C_{\bar l_i h_{\alpha} l_k^-}^RC_{\bar l_k^- h_{\beta} l_j}^RC_{\bar l_i h_{\beta} l_m^-}^LC_{\bar l_m^- h_{\alpha} l_i}^L,\nonumber\\
	&&B_3^L=\frac{m_{\nu_{k}}m_{\nu_{\alpha}}}{e^2m_W^2}G_3(x_{\nu_{k}},x_{\nu_{\alpha}},x_{H^\pm_{\beta}},x_{H^\pm_{l}})[C_{\bar l_i H^\pm_{l} \nu_{k}}^LC_{\bar \nu_{k} H^{\pm*}_{\beta} l_j}^LC_{\bar l_i H^\pm_{\beta} \nu_{\alpha}}^LC_{\bar \nu_{\alpha} H^{\pm*}_{l} l_i}^L\nonumber\\
	&&\qquad\quad-\frac{1}{2}C_{\bar l_i H^\pm_{l} \nu_{k}}^LC_{\bar \nu_{k} H^{\pm*}_{\beta} l_j}^LC_{\bar l_i H^\pm_{l} \nu_{\alpha}}^LC_{\bar \nu_{\alpha} H^{\pm*}_{\beta}l_i}^L]+\frac{m_{l_k^-}m_{l_m^-}}{2e^2m_W^2}G_3(x_{l_k^-},x_{l_m^-},x_{h_{\beta}},x_{h_{\alpha}})\nonumber\\
	&&\qquad\quad\times C_{\bar l_i h_{\alpha} l_k^-}^LC_{\bar l_k^- h_{\beta} l_j}^LC_{\bar l_i h_{\beta} l_m^-}^LC_{\bar l_m^- h_{\alpha} l_i}^L,\nonumber\\
	&&B_4^L=\frac{m_{\nu_{k}}m_{\nu_{\alpha}}}{8e^2m_W^2}G_3(x_{\nu_{k}},x_{\nu_{\alpha}},x_{H^\pm_{\beta}},x_{H^\pm_{l}})C_{\bar l_i H^\pm_{l} \nu_{k}}^LC_{\bar \nu_{k} H^{\pm*}_{\beta} l_j}^LC_{\bar l_i H^\pm_{l} \nu_{\alpha}}^LC_{\bar \nu_{\alpha} H^{\pm*}_{\beta}l_i}^L,\nonumber\\
	&&B_{1,2,3,4}^R=B_{1,2,3,4}^L({L\leftrightarrow R}).
\end{eqnarray}


\newpage

\begin{thebibliography}{99}
\bibitem{MEGII:2025gzr} K.~Afanaciev \textit{et al.} [MEG II], Eur. Phys. J. C \textbf{85}, no.10, 1177 (2025) [erratum: Eur. Phys. J. C \textbf{85}, no.11, 1317 (2025)].
\bibitem{Baldini:2013ke}A. M. Baldini {\it et al.}, arXiv:1301.7225 [physics.ins-det].
\bibitem{Bellgardt:1987du} U. Bellgardt {\it et al.} [SINDRUM Collaboration], Nucl. Phys. B {\bf 299}, 1 (1988).
\bibitem{Blondel:2013ia}A. Blondel {\it et al.}, arXiv:1301.6113 [physics.ins-det].
\bibitem{Aubert:2009ag}B. Aubert {\it et al.} [BaBar Collaboration], Phys. Rev. Lett. {\bf 104}, 021802 (2010) [arXiv:0908.2381 [hep-ex]].
\bibitem{Hayasaka:2013dsa}K. Hayasaka [Belle and Belle-II Collaborations], J. Phys. Conf. Ser. {\bf 408}, 012069 (2013).
\bibitem{Hayasaka:2010np}K. Hayasaka {\it et al.}, Phys. Lett. B {\bf 687}, 139 (2010) [arXiv:1001.3221 [hep-ex]].

\bibitem{Ilakovac:1994kj}A. Ilakovac and A. Pilaftsis, Nucl. Phys. B {\bf 437}, 491 (1995) [hep-ph/9403398].
\bibitem{Diaz:2000cm}R. Diaz, R. Martinez and J. A. Rodriguez, Phys. Rev. D {\bf 63}, 095007 (2001) [hep-ph/0010149].
\bibitem{Kakizaki:2003jk}M. Kakizaki, Y. Ogura and F. Shima, Phys. Lett. B {\bf 566}, 210 (2003) [hep-ph/0304254].
\bibitem{Arganda:2005ji}E. Arganda and M. J. Herrero, Phys. Rev. D {\bf 73}, 055003 (2006) [hep-ph/0510405].
\bibitem{Toma:2013zsa}T. Toma and A. Vicente, JHEP {\bf 1401}, 160 (2014) [arXiv:1312.2840 [hep-ph]].
\bibitem{Zhang:2014osa}H. B. Zhang, T. F. Feng, S. M. Zhao and F. Sun, Int. J. Mod. Phys. A {\bf 29}, 1450123 (2014) [arXiv:1407.7365 [hep-ph]].
\bibitem{Zhao:2015dna}S. M. Zhao, T. F. Feng, H. B. Zhang, X. J. Zhan, Y. J. Zhang and B. Yan, Phys. Rev. D {\bf 92}, 115016 (2015) [arXiv:1507.06732 [hep-ph]].
\bibitem{Aebischer:2019blw}J.~Aebischer, A.~J.~Buras, M.~Cerd{\`a}-Sevilla and F.~De Fazio, JHEP \textbf{02}, 183 (2020).
\bibitem{Colangelo:2025nbd}P.~Colangelo, F.~De Fazio and D.~Milillo, JHEP \textbf{12}, 069 (2025)


\bibitem{Aliberti:2025beg}R.~Aliberti, T.~Aoyama, E.~Balzani, A.~Bashir, G.~Benton, J.~Bijnens, V.~Biloshytskyi, T.~Blum, D.~Boito and M.~Bruno, \textit{et al.}[arXiv:2505.21476 [hep-ph]].
\bibitem{Boccaletti:2024guq}A.~Boccaletti, S.~Borsanyi, A.~Cotellucci, M.~Davier, Z.~Fodor, F.~Frech, A.~Gerardin, D.~Giusti, A.~Y.~Kotov and L.~Lellouch, \textit{et al.} Nature \textbf{653}, no.8114, 373-377 (2026)

\bibitem{Weinberg:1979sa}S.~Weinberg, Phys. Rev. Lett. \textbf{43} (1979), 1566-1570doi:10.1103/PhysRevLett.43.1566
\bibitem{Hisano:1995nq}J.~Hisano, T.~Moroi, K.~Tobe, M.~Yamaguchi and T.~Yanagida, Phys. Lett. B \textbf{357} (1995), 579-587doi:10.1016/0370-2693(95)00954-J[arXiv:hep-ph/9501407 [hep-ph]].
\bibitem{Gell-Mann:1979vob}M.~Gell-Mann, P.~Ramond and R.~Slansky, Conf. Proc. C \textbf{790927} (1979), 315-321[arXiv:1306.4669 [hep-th]].
\bibitem{Mohapatra:1979ia}R.~N.~Mohapatra and G.~Senjanovic, Phys. Rev. Lett. \textbf{44} (1980), 912doi:10.1103/PhysRevLett.44.912
\bibitem{Canetti:2012kh}L.~Canetti, M.~Drewes, T.~Frossard and M.~Shaposhnikov, Phys. Rev. D \textbf{87} (2013), 093006doi:10.1103/PhysRevD.87.093006[arXiv:1208.4607 [hep-ph]].
\bibitem{Abada:2007ux}A.~Abada, C.~Biggio, F.~Bonnet, M.~B.~Gavela and T.~Hambye, JHEP \textbf{12} (2007), 061doi:10.1088/1126-6708/2007/12/061 [arXiv:0707.4058 [hep-ph]].
\bibitem{Yang:2024znv}J.~L.~Yang, H.~B.~Zhang and T.~F.~Feng, Eur. Phys. J. C \textbf{84} (2024) no.6, 616doi:10.1140/epjc/s10052-024-12958-5 [arXiv:2405.17807 [hep-ph]].
\bibitem{Cao:2025zwn}Z.~Cao, Z.~J.~Yang, J.~L.~Yang and T.~F.~Feng, [arXiv:2510.22537 [hep-ph]].
\bibitem{Yang:2024kfs}J.~L.~Yang, H.~B.~Zhang and T.~F.~Feng, Phys. Lett. B \textbf{853} (2024), 138677doi:10.1016/j.physletb.2024.138677 [arXiv:2404.15990 [hep-ph]].
\bibitem{Yang:2024duo}J.~L.~Yang and J.~Li, Phys. Rev. D \textbf{110} (2024) no.11, 115007doi:10.1103/PhysRevD.110.115007 [arXiv:2411.01744 [hep-ph]].
\bibitem{Schwinger:1948iu}J.~S.~Schwinger, Phys. Rev. \textbf{73} (1948), 416-417doi:10.1103/PhysRev.73.416
\bibitem{Bennett:2006fi}G. W. Bennett {\it et al.} [Muon g-2 Collaboration], Phys. Rev. D {\bf 73}, 072003 (2006)[hep-ex/0602035].
\bibitem{Mohr:2008fa}P. J. Mohr, B. N. Taylor and D. B. Newell, Rev. Mod. Phys. {\bf 80}, 633 (2008) [arXiv:0801.0028 [physics.atom-ph]].
\bibitem{Abel:1991dv}S. A. Abel, W. N. Cottingham and I. B. Whittingham, Phys. Lett. B {\bf 259}, 307 (1991).
\bibitem{Moroi:1995yh}T. Moroi, Phys. Rev. D {\bf 53}, 6565 (1996) Erratum: [Phys. Rev. D {\bf 56}, 4424 (1997)] [hep-ph/9512396].
\bibitem{Feng:2001tr}J. L. Feng and K. T. Matchev, Phys. Rev. Lett. {\bf 86}, 3480 (2001) [hep-ph/0102146].
\bibitem{Martin:2001st}S. P. Martin and J. D. Wells, Phys. Rev. D {\bf 64}, 035003 (2001) [hep-ph/0103067].
\bibitem{Diaz:2002tp}R. A. Diaz, hep-ph/0212237.
\bibitem{Cheung:2009fc}K. Cheung, O. C. W. Kong and J. S. Lee, JHEP {\bf 0906}, 020 (2009) [arXiv:0904.4352 [hep-ph]].
\bibitem{Zhao:2014dxa}S. M. Zhao, T. F. Feng, H. B. Zhang, B. Yan and X. J. Zhan, JHEP {\bf 1411}, 119 (2014) [arXiv:1405.7561 [hep-ph]].
\bibitem{Feng:2008cn}T. F. Feng, L. Sun and X. Y. Yang, Nucl. Phys. B {\bf 800}, 221 (2008) [arXiv:0805.1122 [hep-ph]].
\bibitem{Feng:2008nm}T. F. Feng, L. Sun and X. Y. Yang, Phys. Rev. D {\bf 77}, 116008 (2008) [arXiv:0805.0653 [hep-ph]].
\bibitem{Feng:2009gn}T. F. Feng and X. Y. Yang, Nucl. Phys. B {\bf 814}, 101 (2009) [arXiv:0901.1686 [hep-ph]].
\bibitem{Yang:2009zzh}X. Y. Yang and T. F. Feng, Phys. Lett. B {\bf 675}, 43 (2009).
\bibitem{Padley:2015uma}B. P. Padley, K. Sinha and K. Wang, Phys. Rev. D \textbf{92}, 055025 (2015).
\bibitem{Li:2018aov}S. P. Li, X. Q. Li and Y. D. Yang, Phys. Rev. D \textbf{99}, 035010 (2019).
\bibitem{Li:2020dbg}S.~P.~Li, X.~Q.~Li, Y.~Y.~Li, Y.~D.~Yang and X.~Zhang, JHEP \textbf{01}, 034 (2021).
\bibitem{Cao:2021lmj}J.~Cao, Y.~He, J.~Lian, D.~Zhang and P.~Zhu [arXiv:2102.11355 [hep-ph]].
\bibitem{Chen:2021rnl}N.~Chen, B.~Wang and C.~Y.~Yao [arXiv:2102.05619 [hep-ph]].
\bibitem{Yin:2021yqy}W.~Yin [arXiv:2103.14234 [hep-ph]].
\bibitem{Yin:2020afe}W.~Yin and M.~Yamaguchi [arXiv:2012.03928 [hep-ph]].
\bibitem{Sabatta:2019nfg}D.~Sabatta, A.~S.~Cornell, A.~Goyal, M.~Kumar, B.~Mellado and X.~Ruan, Chin. Phys. C \textbf{44}, 063103 (2020).
\bibitem{vonBuddenbrock:2019ajh}S.~Buddenbrock, A.~S.~Cornell, Y.~Fang, A.~Fadol Mohammed, M.~Kumar, B.~Mellado and K.~G.~Tomiwa, JHEP \textbf{10}, 157 (2019).
\bibitem{vonBuddenbrock:2016rmr}S.~von Buddenbrock, N.~Chakrabarty, A.~S.~Cornell, D.~Kar, M.~Kumar, T.~Mandal, B.~Mellado, B.~Mukhopadhyaya, R.~G.~Reed and X.~Ruan, Eur. Phys. J. C \textbf{76}, 580 (2016).
\bibitem{Okada:2016wlm}N.~Okada and H.~M.~Tran, Phys. Rev. D \textbf{94}, 075016 (2016).
\bibitem{Fukuyama:2016mqb}T.~Fukuyama, N.~Okada and H.~M.~Tran, Phys. Lett. B \textbf{767}, 295-302 (2017).
\bibitem{Belanger:2017vpq}G.~B\'elanger, J.~Da Silva and H.~M.~Tran, Phys. Rev. D \textbf{95}, 115017 (2017).
\bibitem{Megias:2017dzd}E.~Megias, M.~Quiros and L.~Salas, JHEP \textbf{05}, 016 (2017).
\bibitem{Tran:2018kxv}H.~M.~Tran and H.~T.~Nguyen, Phys. Rev. D \textbf{99}, 035040 (2019).
\bibitem{g-2muonQCD}E. Chao, Renwick J. Hudspith, Antoine Gerardin, Jeremy R. Green, Harvey B. Meyer, Konstantin Ottnad, arXiv:2104.02632.
\bibitem{g-2muon}A. E. Carcamo Hernandez, Catalina Espinoza, Juan Carlos Gomez-Izquierdo, Myriam Mondragon, arXiv:2104.02730.
\bibitem{g-2muon1}Andreas Crivellin, Martin Hoferichter, arXiv:2104.03202.
\bibitem{g-2muon2}Motoi Endo, Koichi Hamaguchi, Sho Iwamoto, Teppei Kitahara, arXiv:2104.03217.
\bibitem{g-2muon3}Sho Iwamoto, Tsutomu T. Yanagida, Norimi Yokozaki, arXiv:2104.03223.
\bibitem{g-2muon4}Xiao-Fang Han, Tianjun Li, Hong-Xin Wang, Lei Wang, Yang Zhang, arXiv:2104.03227 .
\bibitem{g-2muon5}Giorgio Arcadi, Lorenzo Calibbi, Marco Fedele, Federico Mescia, arXiv:2104.03228.
\bibitem{g-2muon6}Juan C. Criado, Abdelhak Djouadi, Niko Koivunen, Kristjan Muursepp, Martti Raidal, Hardi Veermae, arXiv:2104.03231.
\bibitem{g-2muon7}Bin Zhu, Xuewen Liu, arXiv:2104.03238.
\bibitem{g-2muon8}Yuchao Gu, Ning Liu, Liangliang Su, Daohan Wang, arXiv:2104.03239.
\bibitem{g-2muon9}Hong-Xin Wang, Lei Wang, Yang Zhang, arXiv:2104.03242.
\bibitem{g-2muon10}Melissa van Beekveld, Wim Beenakker, Marrit Schutten, Jeremy de Wit, arXiv:2104.03245.
\bibitem{g-2muon11}Wen Yin, arXiv:2104.03259.
\bibitem{g-2muon12}Fei Wang, Lei Wu, Yang Xiao, Jin Min Yang, Yang Zhang, arXiv:2104.03262.
\bibitem{g-2muon13}Manuel A. Buen-Abad, JiJi Fan, Matthew Reece, Chen Sun, arXiv:2104.03267.
\bibitem{g-2muon14}Pritam Das, Mrinal Kumar Das, Najimuddin Khan, arXiv:2104.03271.
\bibitem{g-2muon15}Murat Abdughani, Yi-Zhong Fan, Lei Feng, Yue-Lin Sming Tsai, Lei Wu, Qiang Yuan, arXiv:2104.03274.
\bibitem{g-2muon16}Chuan-Hung Chen, Cheng-Wei Chiang, Takaaki Nomura, arXiv:2104.03275.
\bibitem{g-2muon17}Shao-Feng Ge, Xiao-Dong Ma, Pedro Pasquini, arXiv:2104.03276.
\bibitem{g-2muon18}M. Cadeddu, N. Cargioli, F. Dordei, C. Giunti, E. Picciau, arXiv:2104.03280.
\bibitem{g-2muon19}Vedran Brdar, Sudip Jana, Jisuke Kubo, Manfred Lindner, arXiv:2104.03282.
\bibitem{g-2muon20}Junjie Cao, Jingwei Lian, Yusi Pan, Di Zhang, Pengxuan Zhu, arXiv:2104.03284.
\bibitem{g-2muon21}Manimala Chakraborti, Sven Heinemeyer, Ipsita Saha, arXiv:2104.03287.
\bibitem{g-2muon22}Masahiro Ibe, Shin Kobayashi, Yuhei Nakayama, Satoshi Shirai, arXiv:2104.03289.
\bibitem{g-2muon23}Peter Cox, Chengcheng Han, Tsutomu T. Yanagida, arXiv:2104.03290.
\bibitem{g-2muon24}K. S. Babu, Sudip Jana, Manfred Lindner, Vishnu P.K., arXiv:2104.03291.
\bibitem{g-2muon25}Chengcheng Han, arXiv:2104.03292.
\bibitem{g-2muon26}Sven Heinemeyer, Essodjolo Kpatcha, Inaki Lara, Daniel E. Lopez-Fogliani, Carlos Munoz, Natsumi Nagata, arXiv:2104.03294.
\bibitem{g-2muon27}Lorenzo Calibbi, M.L. Lopez-Ibanez, Aurora Melis, Oscar Vives, arXiv:2104.03296.
\bibitem{g-2muon28}D.W.P. Amaral, D.G. Cerdeno, A. Cheek, P. Foldenauer, arXiv:2104.03297.
\bibitem{g-2muon29}Yang Bai, Joshua Berger, arXiv:2104.03301.
\bibitem{g-2muon30}Sebastian Baum, Marcela Carena, Nausheen R. Shah, Carlos E. M. Wagner, arXiv:2104.03302.


\bibitem{Hisano:1995cp}J. Hisano, T. Moroi, K. Tobe and M. Yamaguchi, Phys. Rev. D {\bf 53}, 2442 (1996)[hep-ph/9510309].
\bibitem{Huang:2024ozb}Y.~K.~Huang, J.~L.~Yang, S.~K.~Cui and T.~F.~Feng, Phys. Rev. D \textbf{110} (2024) no.7, 075040doi:10.1103/PhysRevD.110.075040 [arXiv:2409.16628 [hep-ph]].
\bibitem{Yang:2018guw}J.~L.~Yang, T.~F.~Feng, Y.~L.~Yan, W.~Li, S.~M.~Zhao and H.~B.~Zhang, Phys. Rev. D \textbf{99} (2019) no.1, 015002doi:10.1103/PhysRevD.99.015002 [arXiv:1812.03860 [hep-ph]].
\bibitem{PDG}F.~Takahashi \textit{et al.} [Particle Data Group], Int. J. Mod. Phys. A \textbf{41}, no.22, 2630011 (2026).


\bibitem{ATLAS:2016cyf} [ATLAS], ATLAS-CONF-2016-045.
\bibitem{Cacciapaglia:2006pk}G.~Cacciapaglia, C.~Csaki, G.~Marandella and A.~Strumia, Phys. Rev. D \textbf{74}, 033011 (2006).
\bibitem{Carena:2004xs} M.~Carena, A.~Daleo, B.~A.~Dobrescu and T.~M.~P.~Tait, Phys. Rev. D \textbf{70}, 093009 (2004)

\bibitem{Zhang:2013hva}H.~B.~Zhang, T.~F.~Feng, S.~M.~Zhao and T.~J.~Gao, Nucl. Phys. B \textbf{873} (2013), 300-324[erratum: Nucl. Phys. B \textbf{879} (2014), 235]doi:10.1016/j.nuclphysb.2013.04.018 [arXiv:1304.6248 [hep-ph]].
\bibitem{Zhang:2013jva}H.~B.~Zhang, T.~F.~Feng, G.~H.~Luo, Z.~F.~Ge and S.~M.~Zhao,JHEP \textbf{07} (2013), 069[erratum: JHEP \textbf{10} (2013), 173]doi:10.1007/JHEP07(2013)069 [arXiv:1305.4352 [hep-ph]].




\end{thebibliography}
\end{document}